\documentclass[
reprint,
superscriptaddress,
amsmath,amssymb,
aps,
pre,
longbibliography,
]{revtex4-2}

\usepackage{graphicx}
\usepackage{xcolor}
\usepackage{dcolumn}
\usepackage{bm}
\usepackage{hyperref}
\usepackage{float}
\usepackage{booktabs}
\usepackage{longtable}
\newcommand\headercell[1]{\smash[b]{\begin{tabular}[t]{@{}c@{}} #1 \end{tabular}}}

\newcommand{\SMSecWListRW}{S1} \newcommand{\SMSecWListRWoffAxis}{S2} \newcommand{\SMSecWListSAWoffAxis}{S3} \newcommand{\SMTabWListRWnmrOneToThree}{S1} \newcommand{\SMTabWListRWnmrSeven}{S5} \newcommand{\SMTabWListRWnmrNine}{S7} \newcommand{\SMTabWListRWnmrEleven}{S9} \newcommand{\SMTabWListRWnmrTwenty}{S18}  \newcommand{\SMTabWListRWnmrsEightSecond}{S20} \newcommand{\SMTabWListRWnmrsTenFirst}{S21} \newcommand{\SMTabWListRWnmrsTenThird}{S23} \newcommand{\SMTabWListRWnr}{S24} \newcommand{\SMTabWListRWnmSixteenToTwentyXOneAAO}{S25}         \newcommand{\SMTabWListSAWnmSixteenToTwentyXTwoAAA}{S34} 

\begin{document}

\title{
Universal negative energetic elasticity in polymer chains:\\
Crossovers among random, self-avoiding, and neighbor-avoiding walks
}

\author{Nobu C.~Shirai}
\thanks{These authors contributed equally}
\email[Corresponding author. ]{shirai@cc.mie-u.ac.jp}
\affiliation{
Center for Information Technologies and Networks, Mie University, Tsu, Mie 514-8507, Japan
}
\author{Naoyuki Sakumichi}
\thanks{These authors contributed equally}
\email[Corresponding author. ]
{sakumichi@gel.t.u-tokyo.ac.jp}
\affiliation{
Faculty of Social Informatics, ZEN University, Shinjuku, Zushi, Kanagawa 249-0007, Japan
}
\affiliation{
Department of Chemistry and Biotechnology, The University of Tokyo, Bunkyo-ku, Tokyo 113-8656, Japan
}

\date{\today}

\begin{abstract}
Negative energetic elasticity in gels challenges the conventional understanding of gel elasticity; despite extensive research, a concise explanation remains elusive. 
In this study, we use the weakly self-avoiding walk (the Domb--Joyce model; DJ model) and interacting self-avoiding walk (ISAW) to investigate the emergence of negative energetic elasticity in polymer chains.
Using exact enumeration, we show that both the DJ model and ISAW exhibit negative energetic elasticity, which is caused by effective soft-repulsive interactions between polymer segments. 
Moreover, we find that a universal scaling law for the internal energy of both models, with a common exponent of $7/4$, holds consistently across both random-walk--self-avoiding-walk and self-avoiding-walk--neighbor-avoiding-walk crossovers.
These findings suggest that negative energetic elasticity is a fundamental and universal property of polymer networks and chains.
\end{abstract}

\maketitle

\section{Introduction}
\label{sec:intro}
Polymer gels are soft solids containing large amounts of solvent.
They are used in the production of everyday items such as jellies and soft contact lenses; for many applications, a particular elastic modulus is required.
Conventionally, the shear modulus $G$ of a gel is predicted using a classical rubber elasticity theory, such as the affine network~\cite{Flory1953book}, phantom network~\cite{JamesGuth1943jul}, or modified phantom network~\cite{ZhongJohnson2016} model.
These models assume that $G$ is predominantly determined by entropic elasticity and therefore is approximately proportional to the absolute temperature ($T$), i.e., $G\approx aT$. 
However, recent studies~\cite{YoshikawaSakai2021,SakumichiSakai2021} discovered cases of negative energetic elasticity, in which $G=aT-b$ with a significantly large negative constant term $-b$ in some narrow range of temperature.
Initially observed in hydrogels~\cite{YoshikawaSakai2021,FujiyabuSakumichi2021,TangChen2023}, negative energetic elasticity was subsequently confirmed in silicone gels~\cite{AoyamaUrayama2023} and calcium carbonate-based gels~\cite{LiuTang2024}. 
These findings revealed a fundamental distinction between gels and rubbers: gels possess an intrinsic solvent-induced negative energetic contribution to the elasticity, whereas rubbers, lacking sufficient solvent content, do not.
Thus, it is necessary to develop novel elastic theories that go beyond the conventional paradigms aligning gel and rubber elasticity.

Various theoretical approaches have been employed to understand the origin of negative energetic elasticity in gels, including lattice models~\cite{ShiraiSakumichi2023,DuarteRizzi2023,IwakiOzaki2024} and all-atom molecular dynamics simulation~\cite{HagitaSakumichi2023}.
Nevertheless, a concise and universally accepted explanation at the microscopic level remains elusive.

\begin{figure}[t!]
\centering
\includegraphics[width=0.99\linewidth]{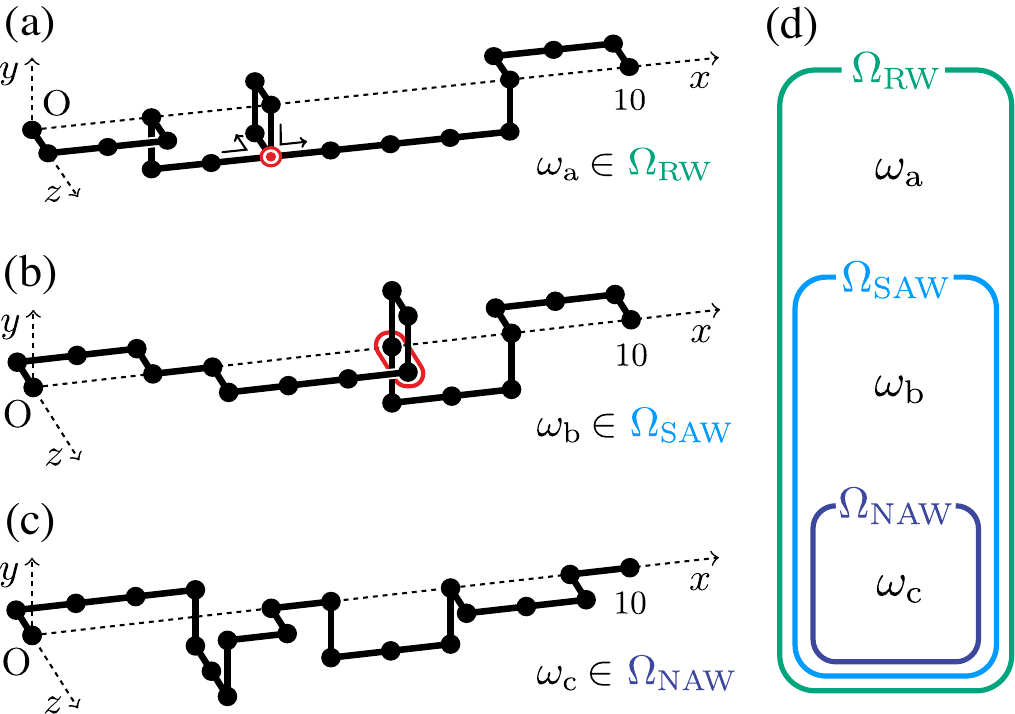}
\caption{
Configurations of 20-step (a) RW $\omega_\mathrm{a}$, (b) SAW $\omega_\mathrm{b}$, and (c) NAW $\omega_\mathrm{c}$ on cubic lattices with endpoints anchored at the origin and $(10,0,0)$.
The red concentric circles in $\omega_\mathrm{a}$ indicate a site occupied by two segments, and the red oval in $\omega_\mathrm{b}$ highlights a pair of nearest-neighbor segments.
(d) Inclusion relations $\Omega_\mathrm{RW} \supset \Omega_\mathrm{SAW} \supset \Omega_\mathrm{NAW}$, where $\Omega_\mathrm{RW}$, $\Omega_\mathrm{SAW}$, and $\Omega_\mathrm{NAW}$ are the configuration spaces of RWs, SAWs, and NAWs, respectively.
}
\label{fig:config}
\end{figure}

In this study, we investigate the microscopic origin of negative energetic elasticity by analyzing the Domb--Joyce (DJ) model~\cite{DombJoyce1972,Domb1983,Slade2019}, a simple extension of the random walk (RW) [Fig.~\ref{fig:config}(a)] with short-range repulsions introduced at the intersections along the polymer chain.
As de Gennes noted~\cite{deGennes1979bookSecI11}, the RW is ``one of the simplest idealizations of a flexible polymer chain'' and serves as a starting point for understanding polymer behavior.
The DJ model reduces to the self-avoiding walk (SAW)~\cite{Orr1947,MadrasSlade1996book} [Fig.~\ref{fig:config}(b)] in the zero-temperature limit; therefore, it can be viewed as a bridge between the RW and SAW, and it has been used to explore the differences in their critical exponents.
Recent studies employed variations of the DJ model (extending from a self-interacting chain to a time series of a non-Markovian walker) to describe agents such as cells, animals, and active matter~\cite{GuerinVoituriez2016,Grassberger2017,dAlessandroLadoux2021,Barbier-ChebbahVoituriez2022,RegnierBenichou2024}.

Through exact enumeration, we demonstrate that the DJ model exhibits negative energetic elasticity and provides a concise and intuitive explanation of its origin in terms of the competition between the energetic and entropic contributions to the elastic response. 
We compare the DJ model with the interacting self-avoiding walk (ISAW)~\cite{Orr1947,Vanderzande1998book} used in Ref.~\cite{ShiraiSakumichi2023} and suggest that any polymer chain possessing effective self-repulsive interactions between its segments exhibits negative energetic elasticity.
Furthermore, we find a universal scaling law for the internal energy of both the DJ model and ISAW, with a common exponent of $7/4$.
These findings suggest that negative energetic elasticity is a universal property of polymer chains and networks, arising from the interplay between the chain's conformational entropy and the effective soft-repulsive interactions between its segments.

The rest of this paper is organized as follows.
In Sec.~\ref{sec:models}, we define and relate five lattice polymer models (RW, SAW, NAW, DJ model, and ISAW) and present the framework to analyze their elastic properties using the finite-difference form of the stiffness.
In Sec.~\ref{sec:exact_enumeration}, we describe our exact enumeration methods for the DJ model and ISAW, validating our results through comparison with existing literature.
In Sec.~\ref{sec:polynomials}, we present polynomials that exactly reproduce the enumeration results for arbitrary chain lengths, enabling analytic calculations of physical quantities.
In Sec.~\ref{sec:emergence}, we demonstrate the emergence of negative energetic elasticity in RW--SAW and SAW--NAW crossovers and offer an intuitive interpretation based on entropic and energetic competition.
In Sec.~\ref{sec:scaling}, we reveal a universal scaling law for the internal energy with an exponent of $7/4$ under the on-axis constraint.
In Sec.~\ref{sec:conclusion}, we summarize our findings and their implications for understanding the elasticity of polymer networks and chains.
Appendixes~\ref{sec:m_upper} to \ref{sec:SAW_polynomials} derive the upper bound of summations and present detailed polynomial expressions, and Appendix~\ref{sec:off-axis_constraints} extends our on-axis scaling analysis to off-axis end-to-end vectors.
All numerical results in this paper are reproducible using the numbers provided in the tables in the Supplemental Material~\cite{SMcommentGelCrossover2023} and Appendix~\ref{sec:SAW_polynomials} together with those in the Supplemental Material of Ref.~\cite{ShiraiSakumichi2023}.

\begin{figure}[t!]
\centering
\includegraphics[width=0.98\linewidth]{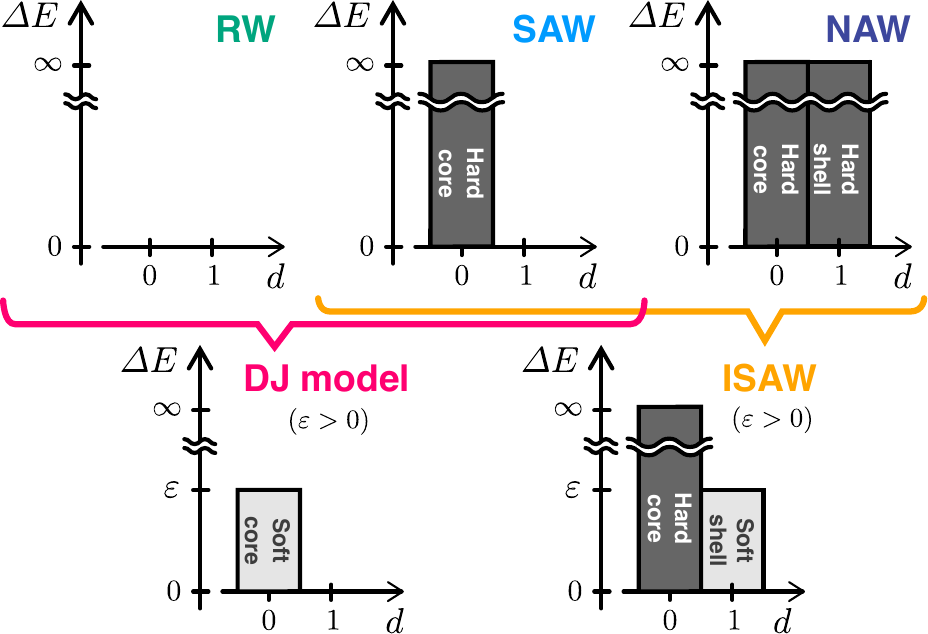}
\caption{
Interaction energy ($\varDelta E$) in five lattice polymer models. 
The range of interactions is denoted by $d$; for $d=0$ or $1$, the interaction is between two segments occupying the same site or nearest-neighbor sites, respectively.
The RW has no interactions between segments.
The SAW introduces a hard-core repulsion ($d=0$) that prohibits segment overlap.
The NAW introduces an additional hard-shell repulsion ($d=1$) between nearest-neighbor segments.
The DJ model and the ISAW incorporate soft-core ($d=0$) and soft-shell ($d=1$) repulsions, respectively, with interaction strength $\varepsilon\,(>0)$.
Varying $\varepsilon$ in these models allows continuous crossovers between the RW, SAW, and NAW. 
}
\label{fig:interaction}
\end{figure}

\section{Models: RW--SAW and SAW--NAW crossovers}
\label{sec:models}
We consider an $n$-step random walk (RW) on a simple cubic lattice with fixed endpoints $\omega(0)=(0,0,0)$ and $\omega(n)=(r_x,r_y,r_z) \in \mathbb{Z}^3$.
We define the end-to-end distance $r=|\omega(n)-\omega(0)|=\sqrt{r_x^2+r_y^2+r_z^2}$.
We mainly focus on the on-axis constraint $\omega(n)=(r,0,0)$ (i.e., $r_y=r_z=0$) as a representative case.
The lattice spacing is set to $1$.
The RW is defined by a sequence of sites $\omega = [\omega(0), \omega(1), \ldots, \omega(n)]$ satisfying $\omega(i) \in \mathbb{Z}^3$ and $|\omega(i + 1) - \omega(i)| = 1$ for $i = 0, 1, \ldots, n - 1$. 
A self-avoiding walk (SAW) is defined as an RW that satisfies $\omega(i) \neq \omega(j)$ for all $i \neq j$~\cite{MadrasSlade1996book}.
Figures~\ref{fig:config}(a) and \ref{fig:config}(b) show configurations of a $20$-step RW ($\omega_\mathrm{a}$) and SAW ($\omega_\mathrm{b}$), respectively.

To investigate the effects of short-range repulsive interactions on the elasticity of the polymer chains, we introduce the DJ model and ISAW as extensions of the RW and SAW, respectively. 
Both models use a unified energy function for the repulsive interaction
\begin{equation}
E(\omega) = \varepsilon\, m(\omega), \label{eq:E}
\end{equation}
where $\varepsilon\,(>0)$ is the repulsive interaction energy (bottom panels of Fig.~\ref{fig:interaction}), $m=m(\omega)$ is the number of interacting segment pairs, and $\omega$ is the configuration of the RW (SAW) for the DJ model (ISAW). 
In the DJ model, $m$ increases whenever multiple segments occupy the same site.
When $v\,(\geq 2)$ segments overlap at one site, $m$ increases by the binomial coefficient $\binom{v}{2}\equiv v(v-1)/2$.
In the ISAW, an interacting segment pair is defined as two sequentially nonadjacent segments occupying nearest-neighbor sites.

Figure~\ref{fig:interaction} illustrates the interactions in five lattice polymer models, revealing the RW--SAW and SAW--NAW crossovers.
The RW [Fig.~\ref{fig:config}(a)] has no interaction between its segments (top left of Fig.~\ref{fig:interaction}).
The SAW [Fig.~\ref{fig:config}(b)] emerges from the RW when a hard-core repulsion is introduced (top center of Fig.~\ref{fig:interaction}) that prohibits segment overlap.
The NAW [Fig.~\ref{fig:config}(c)] emerges from the SAW when an additional hard-shell repulsion is introduced (top right of Fig.~\ref{fig:interaction})~\cite{IshinabeChikahisa1986}. 
The RW, SAW, and NAW all exhibit purely entropic elasticity; energetic elasticity is excluded because all configurations have identical energies.
By contrast, the DJ model and ISAW introduce energy gradients via soft-core and soft-shell repulsions, respectively, with interaction energies $\varepsilon\, (>0)$ (bottom panels of Fig.~\ref{fig:interaction}). 
By varying $\varepsilon$, the DJ model (ISAW) connects the RW (SAW) at $\varepsilon=0$ to the SAW (NAW) at $\varepsilon=\infty$.
These energy gradients enable the DJ model and ISAW to exhibit both energetic and entropic elasticities.

The energy function in Eq.~(\ref{eq:E}) enables the DJ model and ISAW to be treated simultaneously as follows: The partition function under the on-axis constraint is given by
\begin{equation}
Z(r,\beta\varepsilon) = \sum_{m=0}^{m_\mathrm{ub}} W_{n,m}(r)\,e^{- \beta\varepsilon m}, \label{eq:Z}
\end{equation}
where $\beta\,[\equiv 1/(k_\mathrm{B} T)]$ is the inverse temperature, $k_\mathrm{B}$ is the Boltzmann constant, $W_{n,m}(r)$ is the number of possible configurations $\omega$ for given $(n,r,m)$, and $m_\mathrm{ub}$ is an upper bound on $m$ (see Appendix~\ref{sec:m_upper}).
The free energy, internal energy, and entropy are given by
\begin{eqnarray}
A(r,\beta\varepsilon) &=& - \frac{1}{\beta} \ln Z(r,\beta\varepsilon),\\
U(r,\beta\varepsilon) &=& \frac{\varepsilon}{Z(r,\beta\varepsilon)}\, \sum_{m=0}^{m_\mathrm{ub}} m W_{n,m}(r)\,e^{- \beta\varepsilon m},\label{eq:U}\\
S(r,\beta\varepsilon) &=& k_B \beta \left[U(r,\beta\varepsilon)-A(r,\beta\varepsilon)\right],
\end{eqnarray}
respectively.

The finite-difference form of the stiffness for both models is given by
\begin{eqnarray}
k(r,\beta\varepsilon) &\equiv& \frac{1}{\beta}\left[ \left( \frac{1}{Z(r,\beta\varepsilon)}\sum_{m=0}^{m_\mathrm{ub}} \frac{\varDelta W_{n,m}(r)}{\varDelta r} \,e^{-\beta\varepsilon m} \right)^2 \right.\notag\\
&& \left. \quad - \frac{1}{Z(r,\beta\varepsilon)}
\sum_{m=0}^{m_\mathrm{ub}} \frac{\varDelta^2 W_{n,m}(r)}{\varDelta r^2} \,e^{- \beta\varepsilon m}
 \right],
\label{eq:k}
\end{eqnarray}
where $\varDelta W_{n,m}(r) \equiv \left[W_{n,m}(r+\varDelta r)- W_{n,m}(r-\varDelta r)\right]/2$ and $\varDelta^2 W_{n,m}(r) \equiv W_{n,m}(r+\varDelta r) - 2W_{n,m}(r)+ W_{n,m}(r-\varDelta r)$.
Here, $\varDelta r \equiv 2$ because $\omega$ exists only when $r$ and $n$ are both even or both odd.
The stiffness $k$ is the sum of energetic ($k_U$) and entropic ($k_S$) contributions, where $k_U(r,\beta\varepsilon)\equiv \partial^2 U(r,\beta\varepsilon)/\partial r^2$ and $k_S(r,\beta\varepsilon)\equiv -T \partial^2 S(r,\beta\varepsilon)/\partial r^2$.
We use 
\begin{equation}
k_S(r,\beta\varepsilon)= -\beta\varepsilon
\frac{\partial k(r,\beta\varepsilon)}{\partial \beta\varepsilon}, 
\end{equation}
which is derived from Maxwell's relations~\cite{Flory1953book,YoshikawaSakai2021,SakumichiSakai2021,ShiraiSakumichi2023}, and $k_U=k-k_S$.
When it is necessary to distinguish the models, we employ ``RW'' or ``SAW'' as superscripts on $W_{n,m}(r)$, and ``DJ'' and ``ISAW'' on other physical quantities.

\section{Exact enumeration for the DJ model and ISAW}
\label{sec:exact_enumeration}
We exactly enumerated $W^\mathrm{RW}_{n,m}(r)$ for $n=1,\dots,20$, $W^\mathrm{RW}_{n,m}(n-8)$ for $n=21,22,23$, $W^\mathrm{RW}_{n,m}(n-10)$ for $n=21,\dots,26$, and $W^\mathrm{SAW}_{n,m}(n-10)$ for $n=21,\dots,29$, employing the simplest recursive algorithm~\cite{Martin1962} with two pruning algorithms.
Here, we considered the octahedral symmetry of the simple cubic lattice and the reachability of $\omega$ to an endpoint on the $x$ axis (the method is detailed in Sec.~S2 in the Supplemental Material of Ref.~\cite{ShiraiSakumichi2023}).

Comprehensive lists of $W^\mathrm{RW}_{n,m}(r)$ values obtained from the exact enumerations are provided in Sec.~\SMSecWListRW{} of the Supplemental Material~\cite{SMcommentGelCrossover2023}.
Table~\ref{tab:W_n_r_m_n20} provides an illustrative subset of the enumeration results for $W^\mathrm{RW}_{n,m}(r)$.

\begin{table}[t!]
\centering
\caption{\label{tab:W_n_r_m_n20}List of $W^{\mathrm{RW}}_{n,m}(r)$ for $n=20$ and $r=8,10,12$, which is a part of Table~\SMTabWListRWnmrTwenty{} in Supplemental Material~\cite{SMcommentGelCrossover2023}.
}
\begin{tabular}{@{}*{5}{r} @{}}
\hline
\hline
\headercell{\\$m$} & & \multicolumn{3}{c@{}}{$r$}\\
\cmidrule(l){3-5}
 & & $8$ & $10$ & $12$\\
\hline
$0$ & & $14322531084$ & $2625286352$ & $227589504$\\
$1$ & & $43706172200$ & $5625406136$ & $303906776$\\
$2$ & & $63499792684$ & $5648544020$ & $198896588$\\
$3$ & & $55563554760$ & $3369701616$ & $75755328$\\
$4$ & & $39961462284$ & $1872169836$ & $33960789$\\
$5$ & & $25111433136$ & $855424074$ & $10036914$\\
$6$ & & $13673567535$ & $359838406$ & $3635132$\\
$7$ & & $7249781378$ & $161594846$ & $1310080$\\
$8$ & & $4078923694$ & $73451960$ & $466777$\\
$9$ & & $2069333612$ & $32887440$ & $204330$\\
$10$ & & $1139672875$ & $15051476$ & $67964$\\
$11$ & & $509421524$ & $4755638$ & $12658$\\
$12$ & & $296702202$ & $3405656$ & $14197$\\
$13$ & & $187872584$ & $1719114$ & $4520$\\
$14$ & & $111262038$ & $916660$ & $2236$\\
$15$ & & $36803762$ & $135714$ & $0$\\
$16$ & & $26046711$ & $201446$ & $512$\\
$17$ & & $18453292$ & $99040$ & $88$\\
$18$ & & $8264852$ & $28494$ & $0$\\
$19$ & & $5808674$ & $23132$ & $0$\\
$20$ & & $2856074$ & $9904$ & $12$\\
$21$ & & $1912686$ & $6762$ & $0$\\
$22$ & & $1154864$ & $2870$ & $0$\\
$23$ & & $249366$ & $0$ & $0$\\
$24$ & & $217799$ & $180$ & $0$\\
$25$ & & $255924$ & $456$ & $0$\\
$26$ & & $141410$ & $90$ & $0$\\
$27$ & & $46398$ & $0$ & $0$\\
$28$ & & $40100$ & $0$ & $0$\\
$29$ & & $4800$ & $0$ & $0$\\
$30$ & & $11812$ & $10$ & $0$\\
$31$ & & $5552$ & $0$ & $0$\\
$32$ & & $3258$ & $0$ & $0$\\
$33$ & & $140$ & $0$ & $0$\\
$34$ & & $210$ & $0$ & $0$\\
$36$ & & $434$ & $0$ & $0$\\
$37$ & & $84$ & $0$ & $0$\\
$42$ & & $8$ & $0$ & $0$\\
\hline
\hline
\end{tabular}
\end{table}

To validate the exact enumeration results for $W_{n,m}^\mathrm{RW}(r)$ presented in Sec.~\SMSecWListRW{} of the Supplemental Material~\cite{SMcommentGelCrossover2023}, we confirm their consistency with those reported in Refs.~\cite{ShiraiSakumichi2023,Domb1960,DombWilmers1965,Joyce1973,ButeraComi1999,Hollos2007web,A002896,A135390} using two independent approaches.
First, we validate the values of $W_{n,m}^\mathrm{RW}(r)$ for $m=0$ using the equality of the number of $n$-step RWs without intersections $W_{n,0}^\mathrm{RW}(r)$ and the number of $n$-step SAWs $W_n^\mathrm{SAW}(r)$.
We insert our enumerated values of $W_{n,0}^\mathrm{RW}(r)$ into the left side of $W_{n,0}^\mathrm{RW}(r) = W_n^\mathrm{SAW}(r)$, and the reference values of $W_n^\mathrm{SAW}(r)$ from Refs.~\cite{DombWilmers1965,ButeraComi1999} and Table~S1 of Ref.~\cite{ShiraiSakumichi2023} into the right side; the agreement was perfect.

Second, we validate the values of $W_{n,m}^\mathrm{RW}(r)$ using the relation
$W_n^\mathrm{RW}(r) = \sum_{m=0}^{m_\mathrm{ub}} W_{n,m}^\mathrm{RW}(r)$,
where $W_n^\mathrm{RW}(r)$ is the total number of $n$-step RWs in a cubic lattice from the origin to the site $(r,0,0)$. 
To compute the reference values for $W_n^\mathrm{RW}(r)$, we adapt the formula given in Ref.~\cite{Hollos2007web} to our notation, with the endpoints anchored at the origin and $(r,0,0)$:
\begin{equation}
\label{eq:W_RW_n_r_binom}
W_n^\mathrm{RW}(r)=\binom{n}{\frac{n-r}{2}} \sum_{k=0}^{\frac{n-r}{2}} \binom{\frac{n-r}{2}}{k} \binom{\frac{n+r}{2}}{k} \binom{2k}{k}.
\end{equation}
Equation~(\ref{eq:W_RW_n_r_binom}) combines the number of possible step arrangements along each axis using binomial coefficients, which represent the number of possible step choices in each axial direction.
Using Eq.~(\ref{eq:W_RW_n_r_binom}), we computed the reference values for $W_n^\mathrm{RW}(r)$ that are listed in Table~\SMTabWListRWnr{} of the Supplemental Material~\cite{SMcommentGelCrossover2023}.
These reference values are in perfect agreement with the literature values for $s=n$ (i.e., $r=0$ for even $n$ and $r=1$ for odd $n$): with Ref.~\cite{Domb1960} for $n=2,4,\dots,12$; with Refs.~\cite{Joyce1973,A002896} for $n=14$ and $16$; and with Ref.~\cite{A135390} for $n=1,3,\dots,17$.
This consistency across multiple independent sources validates our reference values for $W_n^\mathrm{RW}(r)$.
For all $n$ and $r$ considered, there is perfect agreement between the reference values of $W_n^\mathrm{RW}(r)$ and the summations $\sum_{m=0}^{m_\mathrm{ub}} W_{n,m}^\mathrm{RW}(r)$ computed from our enumeration results in Tables~\SMTabWListRWnmrOneToThree{} to \SMTabWListRWnmrTwenty{} of the Supplemental Material~\cite{SMcommentGelCrossover2023}.
This agreement validates the exact enumeration of $W_{n,m}^\mathrm{RW}(r)$.

\section{Polynomials in arbitrary $n$ of $W^\mathrm{RW}_{n,m}(r)$ and $W^\mathrm{SAW}_{n,m}(r)$} \label{sec:polynomials}
Using the enumerated $W^\mathrm{RW}_{n,m}(r)$ and $W^\mathrm{SAW}_{n,m}(r)$ given in Sec.~\SMSecWListRW{} of the Supplemental Material~\cite{SMcommentGelCrossover2023} and Appendix~\ref{sec:SAW_polynomials}, we derive the polynomials in positive integer $n$ that exactly reproduce the numbers $W^\mathrm{RW}_{n,m}(r)$ and $W^\mathrm{SAW}_{n,m}(r)$ for $r=n$, $n-2$, $n-4$, $n-6$, $n-8$, and $n-10$, as shown in Eqs.~(\ref{eq:poly_W_s0_m0}) to (\ref{eq:poly_W_s10_m}) and (\ref{eq:poly_W_SAW_s0_m0}) to (\ref{eq:poly_W_SAW_s0_m}) in Appendixes~\ref{sec:RW_polynomials} and \ref{sec:SAW_polynomials}.
These polynomials enable the calculation of any physical quantity derived from Eq.~(\ref{eq:Z}) for $0 \leq n-r \leq 10$.

To present examples of the polynomials derived using the exact enumeration results,
we focus on the polynomials of $W^\mathrm{RW}_{n,m}(n-10)$ in $n$ for $m=0$ and $1$, which are given in Eqs.~(\ref{eq:poly_W_s10_m0}) and (\ref{eq:poly_W_s10_m1}) in Appendix~\ref{sec:RW_polynomials}. 
To validate these polynomials, we substitute $n=20$ into Eqs.~(\ref{eq:poly_W_s10_m0}) and (\ref{eq:poly_W_s10_m1}).
This substitution yields the values $2625286352$ and $5625406136$, respectively, which are identical to the $m=0$ and $m=1$ values in the $r=10$ column of Table~\ref{tab:W_n_r_m_n20}.
We can extend this validation to $n=21,\dots,26$ by referencing Tables~\SMTabWListRWnmrsTenFirst{} to \SMTabWListRWnmrsTenThird{} in the Supplemental Material~\cite{SMcommentGelCrossover2023}.
By mathematical induction, it is straightforward to prove that Eqs.~(\ref{eq:poly_W_s10_m0}) and (\ref{eq:poly_W_s10_m1}) hold for all $n \geq 27$.

\begin{figure}[t]
\centering
\includegraphics[width=0.99\linewidth]{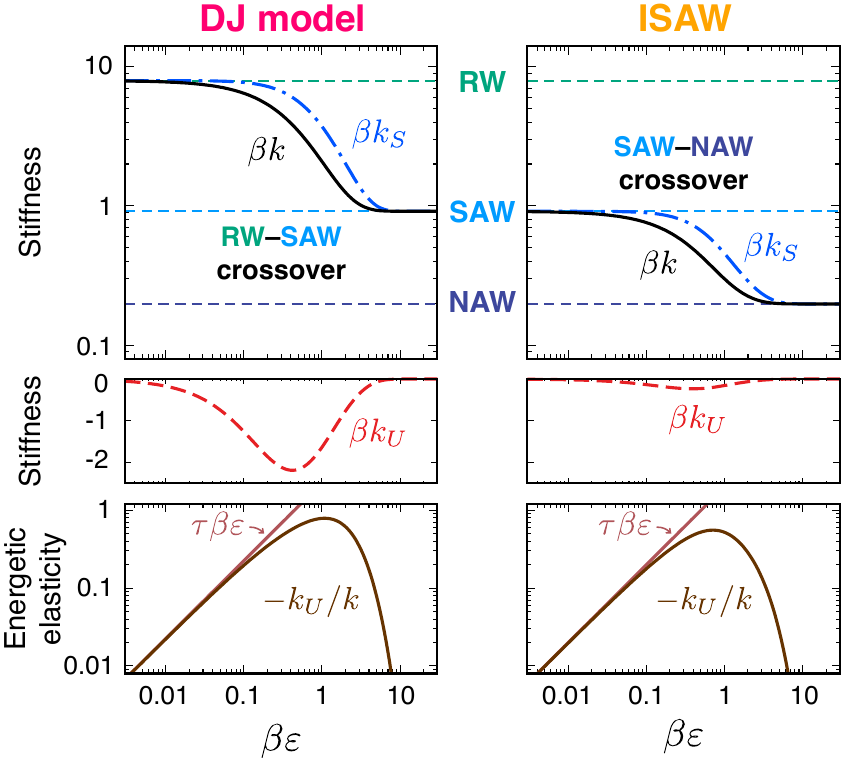}
\caption{
Emergence of negative energetic elasticity in RW--SAW and SAW--NAW crossovers with $(n,r)=(20,10)$.
The DJ model bridges the RW and SAW (left column), and the ISAW bridges the SAW and NAW (right column).
The data for the ISAW are obtained from Ref.~\cite{ShiraiSakumichi2023}.
The top panels show the dependence of total stiffness $\beta k$ and the entropic contribution to stiffness $\beta k_S$ on interaction strength $\beta\varepsilon$.
The middle panels show the energetic contribution to stiffness $\beta k_U$.
The bottom panels show the ratio $-k_U/k$ and its first-order approximation $\tau\beta\varepsilon$ around $\beta\varepsilon=0$.
The maximum of $-k_U/k$ occurs around $\beta\varepsilon\simeq 1$ for both models.
}
\label{fig:crossovers}
\end{figure}

\begin{figure}[t]
\centering
\includegraphics[width=0.99\linewidth]{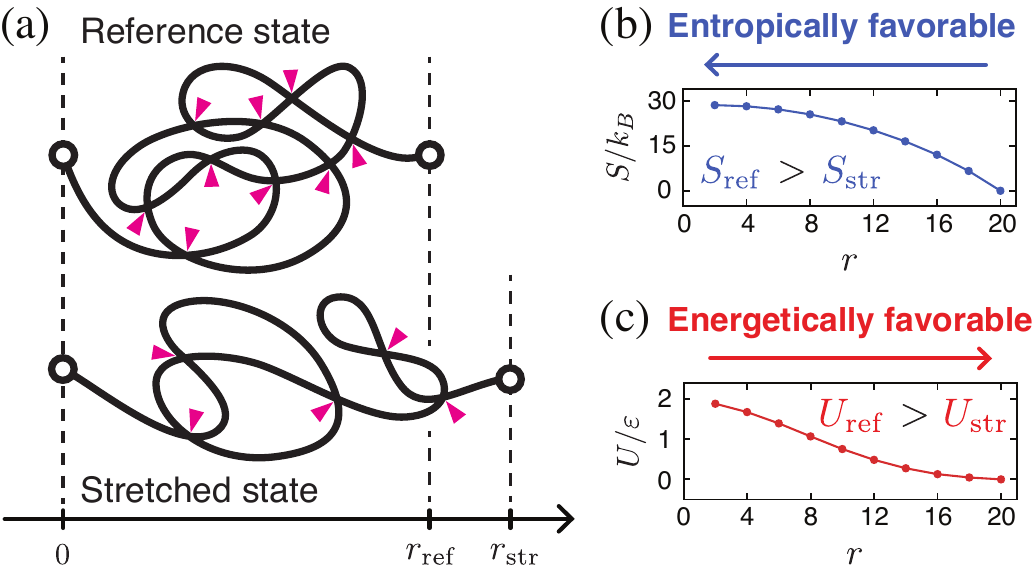}
\caption{
Intuitive interpretation of the emergence of negative energetic elasticity originating from the entropy--energy interplay. 
(a) The reference state with end-to-end distance $r=r_\mathrm{ref}\,(>0)$ is entropically favorable (top), whereas the stretched state with $r=r_\mathrm{str}\,(>r_\mathrm{ref})$ is energetically favorable (bottom) because it has fewer intersections. 
This interpretation is consistent with the behavior of (b) the entropy $S_n^\mathrm{DJ}(r,\beta\varepsilon)$ and (c) the internal energy $U_n^\mathrm{DJ}(r,\beta\varepsilon)$, which are monotonically decreasing functions of $r$ for $1 \leq r\leq 20$.
Here, we set $\beta\varepsilon=1$ and $n=20$.
}
\label{fig:concept}
\end{figure}

\begin{figure*}[t]
\centering
\includegraphics[width=0.99\linewidth]{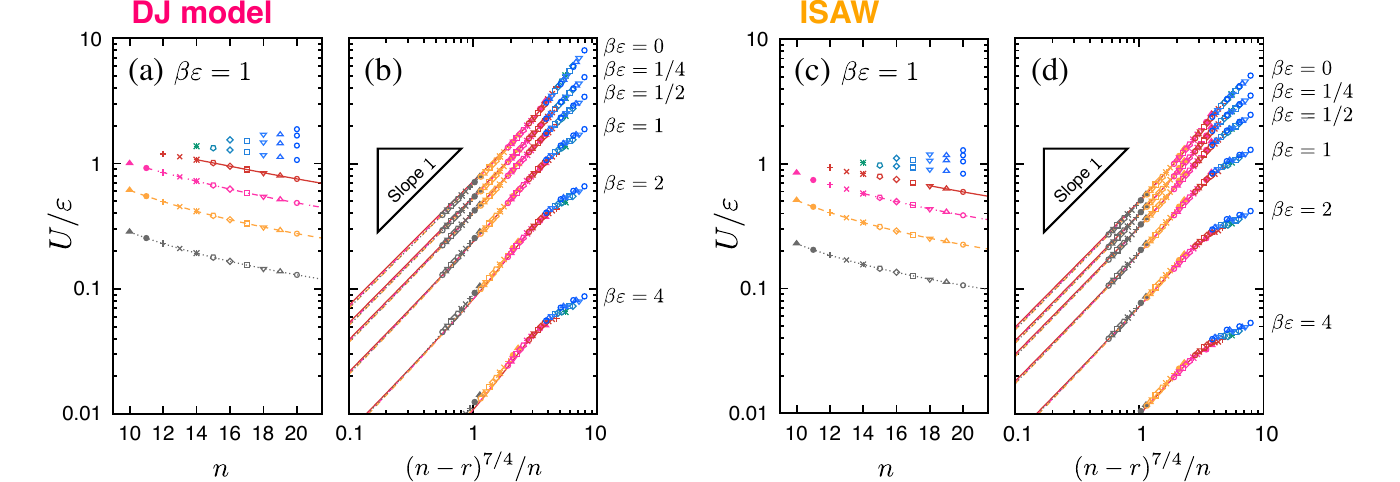}
\caption{
Universal scaling law of the internal energy $U_n(r,\beta\varepsilon)/\varepsilon$ for (a,b) the DJ model across RW--SAW crossover and (c,d) the ISAW across SAW--NAW crossover.
In each panel, points correspond to $n=10,11,\ldots,20$ and $2\leq r \leq n-4$; curves represent the four analytic expressions for $U_n(r,\beta\varepsilon)/\varepsilon$ obtained from the polynomials of $W^\mathrm{RW}_{n,m}(r)$ and $W^\mathrm{SAW}_{n,m}(r)$ (see Appendixes~\ref{sec:RW_polynomials} and \ref{sec:SAW_polynomials}) for $r=n-4$ (gray), $n-6$ (orange), $n-8$ (pink), and $n-10$ (red).
In (a,c), $\beta\varepsilon=1$; for a fixed $n$, $U_n(r,\beta\varepsilon)/\varepsilon$ monotonically decreases with increasing $r$. 
In (b,d), $U_n(r,\beta\varepsilon)/\varepsilon$, plotted against the scaled variable $(n-r)^{7/4}/n$, collapses onto master curves with a common scaling exponent $7/4$ for a wide range of interaction strengths ($\beta\varepsilon=0$, $1/4$, $1/2$, $1$, $2$, and $4$).
The master curves show a approximately linear behavior in the small $(n-r)^{7/4}/n$ regime.
}
\label{fig:U_scaling}
\end{figure*}

\section{Emergence of negative energetic elasticities in lattice polymer chains} \label{sec:emergence}

Figure~\ref{fig:crossovers} shows $\beta k$, $\beta k_S$, $\beta k_U$, $-k_U/k$, and $\tau \beta \varepsilon$ for the DJ model and ISAW for $(n,r)=(20,10)$.
The left column of Fig.~\ref{fig:crossovers} presents these values for the DJ model, calculated using Table~\ref{tab:W_n_r_m_n20}, which enables the computation of 
the differences $\varDelta W_{n,m}(r)$ and $\varDelta^2 W_{n,m}(r)$ at $r=10$ in Eq.~(\ref{eq:k}).
The two crossovers interconnecting the RW, SAW, and NAW ($\beta k_U=0$) reveal that both the DJ model and ISAW exhibit negative energetic elasticity ($\beta k_U<0$) with $|k_U/k|$ maximized around $\beta\varepsilon\sim 1$.
For $(n, r)=(20,10)$ and the same $\beta \varepsilon$, $k^\mathrm{DJ}$ is $4.6$ to $8.7$ times larger than $k^\mathrm{ISAW}$ because the configuration space of the RW is larger than that of the SAW.
The ratio of $|(k_U/k)^\mathrm{DJ}|$ to $|(k_U/k)^\mathrm{ISAW}|$ ranges from $1.1$ to $2.7$, demonstrating the enhanced negative energetic elasticity in the DJ model compared to that in the ISAW.

The soft repulsions between polymer segments in the DJ model and ISAW (Fig.~\ref{fig:interaction}) are the microscopic origin of the negative energetic elasticity observed in Fig.~\ref{fig:crossovers}. 
Our results suggest that any polymer chain, whether alone~\cite{JanshoffFuchs2000,ZhangZhang2003,BaoCui2020} or in a network~\cite{GuJohnson2019,NakagawaYoshie2022}, can exhibit negative energetic elasticity if it possesses effective self-repulsion between polymer segments.

The DJ model suggests an intuitive interpretation of the emergence of negative energetic elasticity in polymer chains.
The stiffness at $r=10$ shown in Fig.~\ref{fig:crossovers} is calculated from the differences in the free energy at $r=8$, $10$, and $12$. 
Therefore, to elucidate the underlying mechanism, we consider the $r$-dependence of both entropy and internal energy.
Figure~\ref{fig:concept}(a) illustrates two configurations of a coarse-grained polymer chain with soft-core repulsion: a reference state and a stretched state.
The energy increases at intersections within the chains [pink triangles in Fig.~\ref{fig:concept}(a)]. 
The stretched state, which possesses fewer intersections, is thus energetically favorable compared to the reference state.
This property holds for the ensemble of chain configurations: states with longer end-to-end distances are energetically favorable; those with shorter end-to-end distances are entropically favorable. 
This behavior is demonstrated by the monotonic decrease of both the entropy $S_n^\mathrm{DJ}(r,\beta\varepsilon)$ [Fig.\ref{fig:concept}(b)] and the internal energy $U_n^\mathrm{DJ}(r,\beta\varepsilon)$ [Fig.~\ref{fig:concept}(c)] with increasing $r$.
This interplay between energetic and entropic factors gives rise to negative energetic elasticity.

The insights from the DJ model can be extended to understand the origin of negative energetic elasticity in gel networks, attributed to the same underlying mechanism.
In Fig. \ref{fig:concept}(a), the polymer chain can be considered a subchain between crosslinks in polymer networks, with $r_\mathrm{ref}$ corresponding to the average distance between crosslinks at the as-prepared state.
Notably, in gels synthesized by end-linking star polymers, where negative energetic elasticity has been observed experimentally~\cite{YoshikawaSakai2021,SakumichiSakai2021,FujiyabuSakumichi2021,TangChen2023}, molecular dynamics simulations~\cite{WangHall2018} and gel fracture experiments~\cite{FujiyabuSakai2022} (see the section ``Failure of Kuhn's model for fracture'' therein) demonstrate that $r_\mathrm{ref}$ is a decreasing function of polymer mass concentration.
At lower concentrations, a larger $r_\mathrm{ref}$ is expected, and the gel becomes unstable below a certain concentration, as confirmed by observations of gel--gel phase separation~\cite{IshikawaSakai2023}.
Since $r_\mathrm{ref}$ depends on concentration, the monotonic decrease of both $S_n^\mathrm{DJ}(r,\beta\varepsilon)$ and $U_n^\mathrm{DJ}(r,\beta\varepsilon)$ with increasing $r$ [Figs.~\ref{fig:concept}(b) and \ref{fig:concept}(c)] becomes crucial for understanding the origin of negative energetic elasticity in gel networks.

\section{Scaling law in internal energy}
\label{sec:scaling}
Figures~\ref{fig:U_scaling}(a) (DJ model) and \ref{fig:U_scaling}(c) (ISAW) show the exact values of $U_n(r,\beta\varepsilon)/\varepsilon$ as a function of the chain length $n$ for $n=10,11,\ldots,20$ at $\beta\varepsilon=1$, 
with the end-to-end distance $2\leq r \leq n-4$.
Also shown are four analytic expressions for $U_n(r,\beta\varepsilon)/\varepsilon$ corresponding to $r=n-4$, $n-6$, $n-8$, and $n-10$.
These analytic expressions for $U_n(r,\beta\varepsilon)/\varepsilon$ are derived by using Eqs.~(\ref{eq:Z}) and (\ref{eq:U}), inserting the polynomials for $W^\mathrm{RW}_{n,m}(r)$ presented in Sec.~\ref{sec:polynomials}.
As an example, we present the analytic expressions of $U^\mathrm{DJ}_n(n-4,\beta\varepsilon)/\varepsilon$ (valid for $n \ge 5$), which corresponds to the gray curves in Figs.~\ref{fig:U_scaling}(a) and \ref{fig:U_scaling}(b):
\begin{equation}
\label{eq:U_DJ_s4}
U^\mathrm{DJ}_n(n-4,\beta\varepsilon)/\varepsilon =\frac{\displaystyle \sum_{m=1}^{6} m\,W^\mathrm{RW}_{n,m}(n-4)\,e^{- \beta\varepsilon m}}{\displaystyle \sum_{m=0}^{6} W^\mathrm{RW}_{n,m}(n-4)\,e^{- \beta\varepsilon m}},
\end{equation}
with the polynomials $W^\mathrm{RW}_{n,m}(n-4)$ given by
\begin{eqnarray*}
W^\mathrm{RW}_{n,0}(n-4) &=&
\frac{1}{2}( 3n^4 -34n^3 +153n^2 -322n +248), \\
W^\mathrm{RW}_{n,1}(n-4) &=&
4(2n^3 -15n^2 +31n -6), \\
W^\mathrm{RW}_{n,2}(n-4) &=&
2n^3 -10n^2 +36n -81, \\
W^\mathrm{RW}_{n,3}(n-4) &=&
4n^2 -22n +50,\\
W^\mathrm{RW}_{n,4}(n-4) &=&
\frac{1}{2}(n^2 +29n -110), \\
W^\mathrm{RW}_{n,5}(n-4) &=&
2(n -5), \\
W^\mathrm{RW}_{n,6}(n-4) &=&
n -4, \\
W^\mathrm{RW}_{n,m}(n-4) &=&
0
\qquad (\textrm{for}\;\; m \ge 7).
\end{eqnarray*}
These polynomials are taken from Eqs.~(\ref{eq:poly_W_s4_m0}) to (\ref{eq:poly_W_s4_m}) in Appendix~\ref{sec:RW_polynomials}.

A remarkable scaling law emerges when the data points and analytic expressions for $U_n(r,\beta\varepsilon)/\varepsilon$ are plotted against the scaled variable $(n-r)^{7/4}/n$. 
For both the DJ model [Fig.~\ref{fig:U_scaling}(b)] and ISAW [Fig.~\ref{fig:U_scaling}(d)], all data points and analytic expressions for $U_n(r,\beta\varepsilon)/\varepsilon$ collapse onto a single master curve across the entire range of $\beta\varepsilon$ ($0$, $1/4$, $1/2$, $1$, $2$, and $4$).
This scaling law persists even for off-axis end-to-end vectors along $(1,1,0)$ and $(1,1,1)$, as shown in Appendix~\ref{sec:off-axis_constraints}.
The component $n-r$ appearing in the scaling variable $(n-r)^{7/4}/n$ is the ``slack'' of the chain, i.e., the number of segments remaining after assigning $r$ segments in the positive $x$ direction to reach the endpoint $(r,0,0)$.
Maclaurin expansions of the analytic expressions for $U_n(n-r,\beta\varepsilon)/\varepsilon$ at $n-r = 4, 6, 8$, and $10$ with respect to $1/n$ yield linear leading-order terms, explaining the observed linear behavior at small $(n - r)^{7/4}/n$.

The emergence of a common exponent $7/4$ in both the DJ model and the ISAW suggests its universality governed by spatial dimensionality, independent of microscopic model details.
Remarkably, this scaling law for the internal energy based on the slack $n-r$ is consistent with the scaling behavior of $\tau^\mathrm{ISAW}$ reported in Ref.~\cite{ShiraiSakumichi2023}, which exhibits an exponent of $3/4$.
Here, 
\begin{equation}
\tau \equiv 
- \frac{\partial}{\partial (\beta\varepsilon)} \left.\left(\frac{k_U}{k}\right)\right|_{\beta\varepsilon = 0}
\end{equation}
is the first-order coefficient in the Maclaurin series of $-k_U/k$ with respect to $\beta\varepsilon$ (see bottom panels of Fig.~\ref{fig:crossovers}).
Reference~\cite{ShiraiSakumichi2023} denotes this coefficient as $T^\infty_U$ instead of $\tau$.

The exponent $3/4$ coincides with the well-established universal critical exponent $\nu=3/4$ of the two-dimensional SAW~\cite{Nienhuis1982,Nienhuis1984}.
This coincidence suggests a possible reduction from three to two dimensions induced by the on-axis constraint on the end-to-end vector (see Sec.~S8 in the Supplemental Material of Ref.~\cite{ShiraiSakumichi2023}).
Furthermore, the simple relationship between the exponents (i.e., $7/4=3/4+1$) can potentially be explained by introducing a scaling function similar to that of the Widom scaling hypothesis~\cite{Widom1965,Cardy1996bookSec34}.
The analysis of the DJ model presented here is expected to advance the long-standing effort, dating back to the 1970s~\cite{DombJoyce1972}, to connect the critical exponents of the three-dimensional RW and SAW~\cite{Grassberger1997}.

We note that this common scaling exponent does not imply that the RW and SAW belong to the same universality class.
RW (Gaussian) and SAW universality classes remain distinct with different critical exponents ($\nu = 0.5$ for RW and $\nu \simeq 0.588$ for SAW in three dimensions) that govern the scaling of spatial extent with chain length.
The $7/4$ exponent identified here relates specifically to the internal energy under stretching, not to the critical exponents that define these universality classes.

\section{Conclusion}
\label{sec:conclusion}
We elucidated the microscopic origin of negative energetic elasticity in polymer chains using the DJ model and ISAW.
Our investigation of the two crossovers among the RW, SAW, and NAW (Fig.~\ref{fig:crossovers}) revealed that effective soft-repulsive interactions between polymer segments are the origin of the negative energetic elasticity of flexible polymers.
The DJ model provides an intuitive interpretation (Fig.~\ref{fig:concept}): longer end-to-end distances are energetically favorable, whereas shorter distances are entropically favorable. 
We also discovered a universal scaling law with an exponent of $7/4$ for the internal energy of both models throughout the crossovers (Fig.~\ref{fig:U_scaling}).
This intuitive interpretation of the negative energetic elasticity, together with the universal scaling law, is expected to serve as a guideline for further theoretical development and material design.

These findings have significant implications for understanding the elasticity of polymer gel networks. 
Our analysis suggests that negative energetic elasticity is a universal property of polymer systems with effective soft-repulsive segment interactions, irrespective of their composition or architecture.
This understanding provides a fundamental framework for explaining the recently discovered negative energetic elasticity in various gels and opens new avenues for designing gel materials with tailored elastic properties.

\begin{acknowledgments}
We thank Kazutaka Takahashi for his insightful comments on the critical exponents.
This study was supported by JSPS KAKENHI Grant No.~JP22K13973 (N.C.S.), No.~JP25K17313 (N.C.S), No.~JP22H01187 (N.S.), No.~JP23K22458 (N.S.), and No.~JP25K00966 (N.S.), and JST FOREST Program  Grant No.~JPMJFR232A (N.S.).
\end{acknowledgments}

\section*{Data Availability}
All exact enumeration data used in this study are fully provided within the Supplemental Material. Additional processed data are not publicly available but are available from the authors upon reasonable request.

\appendix

\section{Upper bounds on the number of interacting segment pairs}
\label{sec:m_upper}

We derive upper bounds on the number of interacting segment pairs $m$ for the DJ model and ISAW.
These upper bounds are used in Eqs.~(\ref{eq:Z}) and (\ref{eq:k}).

For the DJ model, the tightest upper bound on (i.e., the maximum value of) $m$ for a given chain length $n$ and end-to-end distance $r$ is achieved when the overlaps of segments are restricted to the minimum number of sites. 
This is because the number of interacting segment pairs at a single site increases quadratically with the number of overlapping segments $v$ at that site, according to the binomial coefficient $\binom{v}{2} \equiv v(v-1)/2$ for $v \geq 2$.
To construct a configuration that maximizes $m$ for a given $n$ and $r$, we arrange the initial $r$ segments to reach the site $(r,0,0)$, and then arrange the remaining $s \equiv n-r$ segments to oscillate back and forth between two sites along the $x$-axis.
For $r=0$, the oscillation is between the sites $(1,0,0)$ and $(0,0,0)$, and for $r \geq 1$, the oscillation is between $(r-1,0,0)$ and $(r,0,0)$, yielding the maximum of $m$ for given $n$ and $r$:
\begin{equation}
\label{eq:m_ub_RW}
m_\mathrm{ub}^\mathrm{RW} = \begin{cases}
\displaystyle
 \left\lfloor \frac{n^2}{4} \right\rfloor = \left\lfloor \frac{s^2}{4} \right\rfloor & (r=0),\\ \\
\displaystyle
\left\lfloor \frac{(n - r + 1)^2}{4} \right\rfloor = \left\lfloor \frac{(s+ 1)^2}{4} \right\rfloor & (r\geq 1).
\end{cases}
\end{equation}
Here, $\lfloor x \rfloor$ is the floor function, which represents the largest integer less than or equal to $x$. 
Equation~(\ref{eq:m_ub_RW}) gives the maximum of $m$ that satisfies $W^\mathrm{RW}_{n,m}(n-s)\ge 1$ in Tables~\SMTabWListRWnmrOneToThree{} to \SMTabWListRWnmrTwenty{} in the Supplemental Material~\cite{SMcommentGelCrossover2023}.
In the summation of Eq.~(\ref{eq:k}) for the DJ model, $m_\mathrm{ub}^\mathrm{RW}$ depends on $r$ for a fixed $n$.
Thus, the maximum of $m_\mathrm{ub}^\mathrm{RW}$, corresponding to the $r-\varDelta r$ case, should be used.

For the ISAW, an upper bound on $m$ is given by
\begin{equation}
m_\mathrm{ub}^\mathrm{SAW} = 2n-3.
\label{eq:m_ub_SAW}
\end{equation}
The derivation of Eq.~(\ref{eq:m_ub_SAW}) is provided in Sec.~S1 in the Supplemental Material of Ref.~\cite{ShiraiSakumichi2023}.

\begin{widetext}

\section{Polynomials for the number of random walks with small slack}
\label{sec:RW_polynomials}

\begingroup
\allowdisplaybreaks
This Appendix presents the polynomial in positive integer $n$ that exactly reproduces the numbers $W^\mathrm{RW}_{n,m}(r)$ for $r=n$, $n-2$, $n-4$, $n-6$, $n-8$, and $n-10$, and for each nonnegative integer $m$.

For $r=n$, only the fully stretched $\omega$ is allowed, and thus
\begin{eqnarray}
W^\mathrm{RW}_{n,0}(n) &=& 1 \qquad (n \ge 1), \label{eq:poly_W_s0_m0}\\
W^\mathrm{RW}_{n,m}(n) &=& 0 \qquad (n \ge 1 \;\;\textrm{and}\;\; m \ge 1). \label{eq:poly_W_s0_m}
\end{eqnarray}

Assuming that $W^\mathrm{RW}_{n,m}(n-2)$ and $W^\mathrm{RW}_{n,m}(n-4)$ are polynomials in $n$ for each nonnegative integer $m$, we can calculate the coefficients of the polynomials
using the numbers in Tables~\SMTabWListRWnmrOneToThree{} to \SMTabWListRWnmrTwenty{} in the Supplemental Material~\cite{SMcommentGelCrossover2023}. 
The resulting polynomials are as follows:
\begin{eqnarray}
W^\mathrm{RW}_{n,0}(n-2) &=&
2(n^2 -3n +2)
\qquad (n \ge 3), \label{eq:poly_W_s2_m0}\\
W^\mathrm{RW}_{n,1}(n-2) &=&
4n -2
\qquad (n \ge 3), \label{eq:poly_W_s2_m1}\\
W^\mathrm{RW}_{n,2}(n-2) &=&
n -2
\qquad (n \ge 2), \label{eq:poly_W_s2_m2}\\
W^\mathrm{RW}_{n,m}(n-2) &=&
0 \qquad (n \ge 2 \;\;\textrm{and}\;\; m \ge 3), \label{eq:poly_W_s2_m}\\
W^\mathrm{RW}_{n,0}(n-4) &=&
\frac{1}{2}( 3n^4 -34n^3 +153n^2 -322n +248)
\qquad (n \ge 5), 
\label{eq:poly_W_s4_m0}\\
W^\mathrm{RW}_{n,1}(n-4) &=&
4(2n^3 -15n^2 +31n -6)
\qquad (n \ge 4), \label{eq:poly_W_s4_m1}\\
W^\mathrm{RW}_{n,2}(n-4) &=&
2n^3 -10n^2 +36n -81
\qquad (n \ge 5), \label{eq:poly_W_s4_m2}\\
W^\mathrm{RW}_{n,3}(n-4) &=&
4n^2 -22n +50
\qquad (n \ge 5), \label{eq:poly_W_s4_m3}\\
W^\mathrm{RW}_{n,4}(n-4) &=&
\frac{1}{2}(n^2 +29n -110)
\qquad (n \ge 5), \label{eq:poly_W_s4_m4}\\
W^\mathrm{RW}_{n,5}(n-4) &=&
2(n -5)
\qquad (n \ge 5), \label{eq:poly_W_s4_m5}\\
W^\mathrm{RW}_{n,6}(n-4) &=&
n -4
\qquad (n \ge 4), \label{eq:poly_W_s4_m6}\\
W^\mathrm{RW}_{n,m}(n-4) &=&
0
\qquad (n \ge 4 \;\;\textrm{and}\;\; m \ge 7).
\label{eq:poly_W_s4_m}
\end{eqnarray}

The polynomials of $W^\mathrm{RW}_{n,m}(n-6)$ are also calculated using the numbers in Tables~\SMTabWListRWnmrSeven{} to \SMTabWListRWnmrTwenty{} in the Supplemental Material~\cite{SMcommentGelCrossover2023} as
\begin{eqnarray}
W^\mathrm{RW}_{n,0}(n-6) &=&
\frac{1}{9}( 5n^6 -129n^5 +1433n^4 -8745n^3 +30962n^2 -60390n +49320)
\qquad (n \ge 7), \label{eq:poly_W_s6_m0}\\
W^\mathrm{RW}_{n,1}(n-6) &=&
6n^5 -119n^4 +916n^3 -3209n^2 +3926n +1920
\qquad (n \ge 7), \label{eq:poly_W_s6_m1}\\
W^\mathrm{RW}_{n,2}(n-6) &=&
3n^5 -32 n^4 + 129n^3 - 908 n^2 + 6608 n - 15792
\qquad (n \ge 7), \label{eq:poly_W_s6_m2}\\
W^\mathrm{RW}_{n,3}(n-6) &=&
\frac{4}{3}( 6n^4 -91n^3 +606n^2 -2321n +4203)
\qquad (n \ge 7), \label{eq:poly_W_s6_m3}\\
W^\mathrm{RW}_{n,4}(n-6) &=&
n^4 +26n^3 -479n^2 +2659n -4986
\qquad (n \ge 8), \label{eq:poly_W_s6_m4}\\
W^\mathrm{RW}_{n,5}(n-6) &=&
6n^3 -19n^2 -183n +424
\qquad (n \ge 8), \label{eq:poly_W_s6_m5}\\
W^\mathrm{RW}_{n,6}(n-6) &=&
\frac{1}{6}( 13n^3 -39n^2 -394n +2244)
\qquad (n \ge 8), \label{eq:poly_W_s6_m6}\\
W^\mathrm{RW}_{n,7}(n-6) &=&
6n^2 +56n -434
\qquad (n \ge 8), \label{eq:poly_W_s6_m7}\\
W^\mathrm{RW}_{n,8}(n-6) &=&
n^2 +37n -282
\qquad (n \ge 8), \label{eq:poly_W_s6_m8}\\
W^\mathrm{RW}_{n,9}(n-6) &=&
4( 7n -39)
\qquad (n \ge 7), \label{eq:poly_W_s6_m9}\\
W^\mathrm{RW}_{n,10}(n-6) &=&
6(n -7)
\qquad (n \ge 7), \label{eq:poly_W_s6_m10}\\
W^\mathrm{RW}_{n,11}(n-6) &=&
0
\qquad (n \ge 6), \label{eq:poly_W_s6_m11}\\
W^\mathrm{RW}_{n,12}(n-6) &=&
n -6
\qquad (n \ge 6), \label{eq:poly_W_s6_m12}\\
W^\mathrm{RW}_{n,m}(n-6) &=&
0
\qquad (n \ge 6 \;\;\textrm{and}\;\; m \ge 13).
\label{eq:poly_W_s6_m}
\end{eqnarray}

The polynomials of $W^\mathrm{RW}_{n,m}(n-8)$ are also calculated using the numbers in Tables~\SMTabWListRWnmrNine{} to \SMTabWListRWnmrsEightSecond{} in Supplemental Material~\cite{SMcommentGelCrossover2023} as
\begin{eqnarray}
W^\mathrm{RW}_{n,0}(n-8) &=&
\frac{1}{288}( 35n^8 -1620n^7 +33606n^6 -407160n^5 +3151827n^4 -15991140n^3 +52024708n^2\notag\\
&& \quad -99143952n +83572992)
\qquad (n \ge 9), \label{eq:poly_W_s8_m0}\\
W^\mathrm{RW}_{n,1}(n-8) &=&
\frac{2}{9}( 10n^7 -383n^6 +6235n^5 -54875n^4 +270421n^3 -654914n^2 +234858n \notag\\
&& +1480932)
\qquad (n \ge 10), \label{eq:poly_W_s8_m1}\\
W^\mathrm{RW}_{n,2}(n-8) &=&
\frac{1}{18}( 10n^7 -182n^6 +466n^5 +5671n^4 +53494n^3 -1346021n^2 +7741074n \notag\\
&& -14786496)
\qquad (n \ge 11), \label{eq:poly_W_s8_m2}\\
W^\mathrm{RW}_{n,3}(n-8) &=&
\frac{1}{3}( 18n^6 -527n^5 +6921n^4 -56627n^3 +321189n^2 -1150790n +1862184)
\quad (n \ge 10), \label{eq:poly_W_s8_m3}\\
W^\mathrm{RW}_{n,4}(n-8) &=&
\frac{1}{12}( 9n^6 +207n^5 -10249n^4 +142181n^3 -987656n^2 +3600772n -5431572)
\quad (n \ge 11), \label{eq:poly_W_s8_m4}\\
W^\mathrm{RW}_{n,5}(n-8) &=&
\frac{1}{3}(21n^5 -169n^4 -3139n^3 +47077n^2 -198602n +203982)
\qquad (n \ge 10), \label{eq:poly_W_s8_m5}\\
W^\mathrm{RW}_{n,6}(n-8) &=&
\frac{1}{6}( 11n^5 -6n^4 -2791n^3 +32559n^2 -179599n +467172)
\qquad (n \ge 11), \label{eq:poly_W_s8_m6}\\
W^\mathrm{RW}_{n,7}(n-8) &=&
\frac{1}{3}( 38n^4 +11n^3 -9242n^2 +84463n -230220)
\qquad (n \ge 11), \label{eq:poly_W_s8_m7}\\
W^\mathrm{RW}_{n,8}(n-8) &=&
\frac{1}{24}( 49n^4 +1630n^3 -32353n^2 +184586n -427872)
\qquad (n \ge 11), \label{eq:poly_W_s8_m8}\\
W^\mathrm{RW}_{n,9}(n-8) &=&
61n^3 -906n^2 +4157n -4410
\qquad (n \ge 11), \label{eq:poly_W_s8_m9}\\
W^\mathrm{RW}_{n,10}(n-8) &=&
\frac{1}{2}(25n^3 -189n^2 +1006n - 8592)
\qquad (n \ge 11), \label{eq:poly_W_s8_m10}\\
W^\mathrm{RW}_{n,11}(n-8) &=&
54n^2 -514n +1338
\qquad (n \ge 11), \label{eq:poly_W_s8_m11}\\
W^\mathrm{RW}_{n,12}(n-8) &=&
4n^3 -71n^2 +1809n - 11386
\qquad (n \ge 10), \label{eq:poly_W_s8_m12}\\
W^\mathrm{RW}_{n,13}(n-8) &=&
4n^2 +258n -2240
\qquad (n \ge 10), \label{eq:poly_W_s8_m13}\\
W^\mathrm{RW}_{n,14}(n-8) &=&
n^2 +166n -1484
\qquad (n \ge 10), \label{eq:poly_W_s8_m14}\\
W^\mathrm{RW}_{n,15}(n-8) &=&
0
\qquad (n \ge 8), \label{eq:poly_W_s8_m15}\\
W^\mathrm{RW}_{n,16}(n-8) &=&
42n -328
\qquad (n \ge 9), \label{eq:poly_W_s8_m16}\\
W^\mathrm{RW}_{n,17}(n-8) &=&
8(n -9)
\qquad (n \ge 9), \label{eq:poly_W_s8_m17}\\
W^\mathrm{RW}_{n,18}(n-8) &=&
0
\qquad (n \ge 8), \label{eq:poly_W_s8_m18}\\
W^\mathrm{RW}_{n,19}(n-8) &=&
0
\qquad (n \ge 8), \label{eq:poly_W_s8_m19}\\
W^\mathrm{RW}_{n,20}(n-8) &=&
n -8
\qquad (n \ge 8), \label{eq:poly_W_s8_m20}\\
W^\mathrm{RW}_{n,m}(n-8) &=& 0
\qquad (n \ge 8 \;\;\textrm{and}\;\; m \ge 21). 
\label{eq:poly_W_s8_m}
\end{eqnarray}

The polynomials of $W^\mathrm{RW}_{n,m}(n-10)$ are also calculated using the numbers in Tables~\SMTabWListRWnmrEleven{} to \SMTabWListRWnmrTwenty{} and \SMTabWListRWnmrsTenFirst{} to \SMTabWListRWnmrsTenThird{} in the Supplemental Material~\cite{SMcommentGelCrossover2023} as
\begin{eqnarray}
W^\mathrm{RW}_{n,0}(n-10) &=&
\frac{1}{3600}( 63n^{10} -4585n^9 +153060n^8 -3081950n^7 +41448599n^6 -389383445n^5 +2591570990n^4 \notag\\
&& \quad -12085704100n^3 +37840144088n^2 -71829549120n +62526614400)
\qquad (n \ge 11), \label{eq:poly_W_s10_m0}\\
W^\mathrm{RW}_{n,1}(n-10) &=&
\frac{1}{144}( 70n^9 -4395n^8 +122592n^7 -1976910n^6 +20015526n^5 -127970403n^4 +476189444n^3 \notag\\
&& \quad -685627428n^2 -1416480720n +5184797184)
\qquad (n \ge 12), \label{eq:poly_W_s10_m1}\\
W^\mathrm{RW}_{n,2}(n-10) &=&
\frac{1}{288}( 35n^9 -970n^8 -1738n^7 +356852n^6 -4644373n^5 +7548302n^4 +338801436n^3 \notag\\
&& \quad -3500058552n^2 +14604644256n -23322994560)
\qquad (n \ge 13), \label{eq:poly_W_s10_m2}\\
W^\mathrm{RW}_{n,3}(n-10) &=&
\frac{2}{9}( 10n^8 -481n^7 +10542n^6 -145552n^5 +1475385n^4 -11407483n^3 +62293463n^2 \notag\\
&& \quad -204521796n +293371992) \qquad (n \ge 11), \label{eq:poly_W_s10_m3}\\
W^\mathrm{RW}_{n,4}(n-10) &=&
\frac{1}{18}( 5n^8 +112n^7 -11966n^6 +304687n^5 -4138721n^4 +34852933n^3 -185222134n^2 \notag\\
&& \quad +568825140n -751096152)
\qquad (n \ge 13), \label{eq:poly_W_s10_m4}\\
W^\mathrm{RW}_{n,5}(n-10) &=&
\frac{1}{90}( 370n^7 -7025n^6 -90077n^5 +3908575n^4 -47417645n^3 +270658210n^2 -667541568n \notag\\
&&\quad +291274020)
\qquad (n \ge 13), \label{eq:poly_W_s10_m5}\\
W^\mathrm{RW}_{n,6}(n-10) &=&
\frac{1}{36}(29n^7 +163n^6 -28069n^5 +546697n^4 -5586400n^3 +39828436n^2 -198287652n \notag\\
&& \quad +477041256)
\qquad (n \ge 13), \label{eq:poly_W_s10_m6}\\
W^\mathrm{RW}_{n,7}(n-10) &=&
\frac{1}{3}( 31n^6 -332n^5 -13194n^4 +342222n^3 -3412555n^2 +16788098n \notag\\
&& \quad -34291422)
\qquad (n \ge 13), \label{eq:poly_W_s10_m7}\\
W^\mathrm{RW}_{n,8}(n-10) &=&
\frac{1}{12}( 19n^6 +601n^5 -24387n^4 +243417n^3 -732250n^2 -572620n +1159920)
\qquad (n \ge 14), \label{eq:poly_W_s10_m8}\\
W^\mathrm{RW}_{n,9}(n-10) &=&
\frac{1}{12}( 626n^5 - 16317n^4 + 148932n^3 - 413523n^2 - 1757150n + 11265480)
\qquad (n \ge 14), \label{eq:poly_W_s10_m9}\\
W^\mathrm{RW}_{n,10}(n-10) &=&
\frac{1}{120}( 1201n^5 - 5560n^4 - 242445n^3 + 1552240n^2 + 14263964n - 114038160)
\qquad (n \ge 14), \label{eq:poly_W_s10_m10}\\
W^\mathrm{RW}_{n,11}(n-10) &=&
\frac{1}{3}( 331n^4 -7697n^3 +78983n^2 -519949n +1688694)
\qquad (n \ge 14), \label{eq:poly_W_s10_m11}\\
W^\mathrm{RW}_{n,12}(n-10) &=&
\frac{1}{6}( 9n^5 -293n^4 +13253n^3 -255709n^2 +2019160n -5609664)
\qquad (n \ge 14), \label{eq:poly_W_s10_m12}\\
W^\mathrm{RW}_{n,13}(n-10) &=&
8n^4 + 406n^3 - 10663n^2 + 86579n - 275266
\qquad (n \ge 14), \label{eq:poly_W_s10_m13}\\
W^\mathrm{RW}_{n,14}(n-10) &=&
\frac{1}{4}( 8n^4 + 1222n^3 - 28774n^2 + 251988n - 919520)
\qquad (n \ge 14), \label{eq:poly_W_s10_m14}\\
W^\mathrm{RW}_{n,15}(n-10) &=&
4n^3 + 974n^2 - 18158n + 77274
\qquad (n \ge 12), \label{eq:poly_W_s10_m15}\\
W^\mathrm{RW}_{n,16}(n-10) &=&
\frac{1}{2}( 169n^3 - 4121n^2 + 44276n - 186228)
\qquad (n \ge 13), \label{eq:poly_W_s10_m16}\\
W^\mathrm{RW}_{n,17}(n-10) &=&
16n^3 -278n^2 +6858n -54920
\qquad (n \ge 13), \label{eq:poly_W_s10_m17}\\
W^\mathrm{RW}_{n,18}(n-10) &=&
75n^2 +538n -12266
\qquad (n \ge 12), \label{eq:poly_W_s10_m18}\\
W^\mathrm{RW}_{n,19}(n-10) &=&
4(2n^2 +512n -5257)
\qquad (n \ge 12), \label{eq:poly_W_s10_m19}\\
W^\mathrm{RW}_{n,20}(n-10) &=&
2(n^3 -27n^2 +614n -4528)
\qquad (n \ge 12), \label{eq:poly_W_s10_m20}\\
W^\mathrm{RW}_{n,21}(n-10) &=&
4n^2 +546n -5758
\qquad (n \ge 12), \label{eq:poly_W_s10_m21}\\
W^\mathrm{RW}_{n,22}(n-10) &=&
n^2 +273n -2990
\qquad (n \ge 12), \label{eq:poly_W_s10_m22}\\
W^\mathrm{RW}_{n,23}(n-10) &=&
0
\qquad (n \ge 10), \label{eq:poly_W_s10_m23}\\
W^\mathrm{RW}_{n,24}(n-10) &=&
20(n -11)
\qquad (n \ge 11), \label{eq:poly_W_s10_m24}\\
W^\mathrm{RW}_{n,25}(n-10) &=&
44n -424
\qquad (n \ge 11), \label{eq:poly_W_s10_m25}\\
W^\mathrm{RW}_{n,26}(n-10) &=&
10(n -11)
\qquad (n \ge 11), \label{eq:poly_W_s10_m26}\\
W^\mathrm{RW}_{n,27}(n-10) &=&
0
\qquad (n \ge 10), \label{eq:poly_W_s10_m27}\\
W^\mathrm{RW}_{n,28}(n-10) &=&
0
\qquad (n \ge 10), \label{eq:poly_W_s10_m28}\\
W^\mathrm{RW}_{n,29}(n-10) &=&
0
\qquad (n \ge 10), \label{eq:poly_W_s10_m29}\\
W^\mathrm{RW}_{n,30}(n-10) &=&
n -10
\qquad (n \ge 10), \label{eq:poly_W_s10_m30}\\
W^\mathrm{RW}_{n,m}(n-10) &=&
0
\qquad (n \ge 10 \;\;\textrm{and}\;\; m \ge 31).
\label{eq:poly_W_s10_m}
\end{eqnarray}

\section{Polynomials for the number of self-avoiding walks with small slack}
\label{sec:SAW_polynomials}

This Appendix presents the polynomial for the SAW in positive integer $n$ that exactly reproduces the numbers $W^\mathrm{SAW}_{n,m}(n-10)$ for each nonnegative integer $m$.
These polynomials extend the polynomials for $W^\mathrm{SAW}_{n,m}(n-2)$, $W^\mathrm{SAW}_{n,m}(n-4)$, $W^\mathrm{SAW}_{n,m}(n-6)$, and $W^\mathrm{SAW}_{n,m}(n-8)$ presented in Sec.~S12 in the Supplemental Material of Ref.~\cite{ShiraiSakumichi2023}, where these quantities are denoted as $W_{n,m}(n-s)$ without the SAW superscript. 

Assuming that $W^\mathrm{SAW}_{n,m}(n-10)$ is a polynomial in $n$ for each nonnegative integer $m$, we can calculate the coefficients of the polynomial using the numbers in Tables~S3 to S19 in the Supplemental Material of Ref.~\cite{ShiraiSakumichi2023}, along with additional numbers for $n=21,\dots,29$ in Table~\ref{tab:W_SAW_n_m_s10_n21to29}.

\begin{table}[H]
\centering
\caption{\label{tab:W_SAW_n_m_s10_n21to29}List of $W^\mathrm{SAW}_{n,m}(n-10)$ for $n=21,\dots,29$.}
\begin{tabular}{@{}*{13}{r} @{}}
\hline
\hline
\headercell{\\$n$} & \multicolumn{11}{c@{}}{$m$}\\
\cmidrule(l){3-13}
 & & $0$ & $1$ & $2$ & $3$ & $4$ & $5$ & $6$ & $7$ & $8$ & $9$ & $10$\\
\hline
$21$ & & $1164644956$ & $1790415816$ & $1381489928$ & $703113644$ & $243996580$ & $60732044$ & $12065116$ & $2130064$ & $263300$ & $27212$ & $3168$\\
$22$ & & $2458183236$ & $3608669000$ & $2631650372$ & $1249816464$ & $400672608$ & $91830736$ & $16926932$ & $2800736$ & $319512$ & $32096$ & $3520$\\
$23$ & & $4949794148$ & $6947511200$ & $4797886376$ & $2132870620$ & $634321436$ & $134522404$ & $23126884$ & $3600992$ & $381248$ & $37332$ & $3872$\\
$24$ & & $9556918148$ & $12842271216$ & $8415109804$ & $3512153384$ & $972785768$ & $191749776$ & $30888004$ & $4542352$ & $448508$ & $42920$ & $4224$\\
$25$ & & $17771135392$ & $22892935256$ & $14261282728$ & $5604132844$ & $1450869908$ & $266904588$ & $40450044$ & $5636336$ & $521292$ & $48860$ & $4576$\\
$26$ & & $31947019844$ & $39504999416$ & $23440268068$ & $8695905504$ & $2111409280$ & $363857968$ & $52069476$ & $6894464$ & $599600$ & $55152$ & $4928$\\
$27$ & & $55705056944$ & $66207779080$ & $37484627332$ & $13161411164$ & $3006415548$ & $486990820$ & $66019492$ & $8328256$ & $683432$ & $61796$ & $5280$\\
$28$ & & $94482703636$ & $108067854720$ & $58481393516$ & $19479984120$ & $4198297336$ & $641224208$ & $82590004$ & $9949232$ & $772788$ & $68792$ & $5632$\\
$29$ & & $156274920996$ & $172217889976$ & $89224065624$ & $28257399724$ & $5761156548$ & $832049740$ & $102087644$ & $11768912$ & $867668$ & $76140$ & $5984$\\
\hline
\hline
\end{tabular}
\end{table}

The resulting polynomials are:
\begin{eqnarray}
W^\mathrm{SAW}_{n,0}(n-10) &=& \frac{1}{3600}(63 n^{10}-6335 n^9+299060 n^8-8750150 n^7+176109699 n^6-2551326815 n^5+26957728290 n^4 \notag\\
&&-205046147100 n^3 +1073071273288 n^2 -3481733392800 n +5303917051200) \qquad (n\ge 18), \label{eq:poly_W_SAW_s0_m0}\\
W^\mathrm{SAW}_{n,1}(n-10) &=& \frac{1}{72}(35 n^9 -3310 n^8 +144274 n^7 -3810720 n^6 +67350079 n^5 -827584026 n^4 +7081605356 n^3 \notag\\
&&-40733780024 n^2 -232990627968 +142946744256 n) \qquad (n\ge 18),\\
W^\mathrm{SAW}_{n,2}(n-10) &=& \frac{1}{36}(195 n^8 -16591 n^7 +636953 n^6 -14439205 n^5 +211951772 n^4 -2069561980 n^3 +13172823368 n^2 \notag\\
&& -50135482272 n +87570505008) \qquad (n\ge 18),\\
W^\mathrm{SAW}_{n,3}(n-10) &=& \frac{4}{9}(71n^7 - 5124n^6 + 161663n^5 - 2893446n^4 + 31815851n^3 - 216064455n^2 + 846392625n \notag\\
&& \quad - 1493841726) \qquad (n\ge 18),\\
W^\mathrm{SAW}_{n,4}(n-10) &=& \frac{1}{3}(315n^6 - 17677n^5 + 401955n^4 - 4587843n^3 + 25622178n^2 - 48389612n\notag\\
&& - 57479976) \qquad (n\ge 18),\\
W^\mathrm{SAW}_{n,5}(n-10) &=& \frac{2}{15}(1899n^5 - 78070n^4 + 1086275n^3 - 3841550n^2 - 34041704n + 231920760) \qquad (n\ge 18),\\
W^\mathrm{SAW}_{n,6}(n-10) &=& \frac{2}{3}(1045n^4 - 38292n^3 + 495149n^2 - 2452554n + 2630166) \qquad (n\ge 18),\\
W^\mathrm{SAW}_{n,7}(n-10) &=& 8(240n^3 - 7741n^2 + 83817n - 302758) \qquad (n\ge 18),\\
W^\mathrm{SAW}_{n,8}(n-10) &=& 
2(1381n^2 - 31277n + 179446) \qquad (n\ge 18),\\
W^\mathrm{SAW}_{n,9}(n-10) &=& 4(44n^2 - 671n + 1490) \qquad (n\ge 18),\\
W^\mathrm{SAW}_{n,10}(n-10) &=& 352(n - 12) \qquad (n\ge 13),\\
W^\mathrm{SAW}_{n,m}(n-10) &=& 0 \qquad (n \ge 10 \;\;\textrm{and}\;\; m\ge 11). \label{eq:poly_W_SAW_s0_m}
\end{eqnarray}

\endgroup

\end{widetext}

\section{Universality of scaling in internal energy for off-axis directions}
\label{sec:off-axis_constraints}

In this Appendix, we extend the analysis of the scaling law for the internal energy presented in Sec.~\ref{sec:scaling} from on-axis to off-axis constraints on the end-to-end vector to demonstrate that the scaling behavior is independent of the direction of the end-to-end vector.
We consider RW and SAW with end-to-end vectors $\mathbf{r}\equiv \omega(n)-\omega(0)$ parallel to $(1,1,0)$ or $(1,1,1)$, i.e., $\mathbf{r}\parallel (1,1,0)$ or $\mathbf{r}\parallel (1,1,1)$.
Specifically, we exactly enumerate $W_{n,m}^\mathrm{RW}(\mathbf{r})$ and $W_{n,m}^\mathrm{SAW}(\mathbf{r})$ for $\mathbf{r}=(1,1,0)$, $(2,2,0)$, $(3,3,0)$, $(1,1,1)$, and $(2,2,2)$ with $15\leq n \leq 20$.
The full enumeration results are provided in Tables~\SMTabWListRWnmSixteenToTwentyXOneAAO{} to \SMTabWListSAWnmSixteenToTwentyXTwoAAA{} of Secs.~\SMSecWListRWoffAxis{} and \SMSecWListSAWoffAxis{} in the Supplemental Material~\cite{SMcommentGelCrossover2023}.

To validate the enumerated values under off-axis constraints, we employ the formula from Ref.~\cite{Hollos2007web}, which gives the number of $n$-step random walks on a cubic lattice with a specified end-to-end vector $\mathbf{r}=(r_x,r_y,r_z)$, following the approach in Sec.~\ref{sec:exact_enumeration}:
\begin{widetext}
\begin{equation}
W_n^{\mathrm{RW}}(\mathbf{r})
=\binom{n}{\frac{n-(r_x+r_y+r_z)}{2}}
\sum_{k=0}^{\frac{n-(r_x+r_y+r_z)}{2}}
\binom{\frac{n-(r_x+r_y+r_z)}{2}}{k}
\binom{\frac{n+r_x+r_y+r_z}{2}}{k + r_y + r_z}
\binom{2k + r_y + r_z}{k + r_z}.
\label{eq:W_RW_n_r_binom_abc}
\end{equation}
\end{widetext}
For each $\mathbf{r}$, the total $\sum_{m}W_{n,m}^\mathrm{RW}(\mathbf{r})$ obtained from our enumeration agrees exactly with $W_n^\mathrm{RW}(\mathbf{r})$ computed via Eq.~(\ref{eq:W_RW_n_r_binom_abc}), verifying our enumeration results under off-axis constraints.

\begin{figure}[t]
\centering
\includegraphics[width=0.99\linewidth]{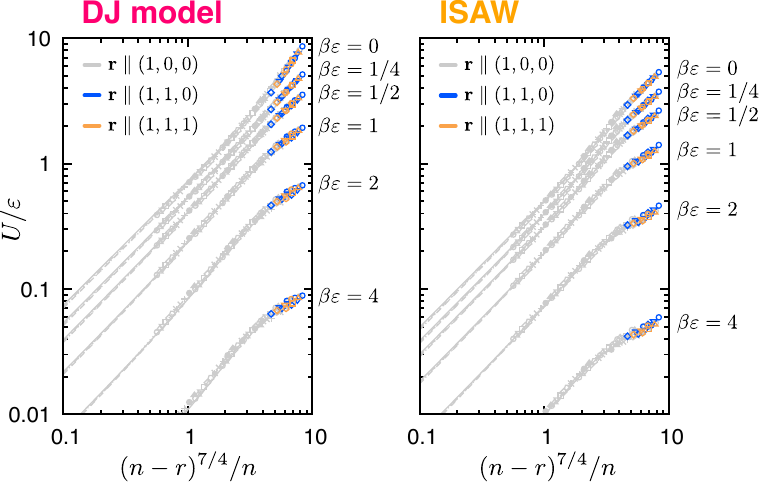}
\caption{
Collapse of $U_n(r,\beta\varepsilon)/\varepsilon$ for off-axis end-to-end constraints (blue and orange) onto the on-axis results (gray, data from Fig.~\ref{fig:U_scaling}). 
The agreement demonstrates that the $7/4$ scaling exponent holds regardless of the direction of the end-to-end vector.
}
\label{fig:U_scaling_off_axis}
\end{figure}

Figure~\ref{fig:U_scaling_off_axis} shows that $U_n(r,\beta\varepsilon)/\varepsilon$ for off-axis end-to-end constraints [along the $(1,1,0)$ and $(1,1,1)$] collapses onto the same $(n - r)^{7/4}/n$ master curve observed for the on-axis constraint (Fig.~\ref{fig:U_scaling}).
This confirms that the $7/4$ scaling exponent is independent of the direction of the end-to-end vector.

\end{document}


\title{
Supplemental Material for:\\
``Universal negative energetic elasticity in polymer chains:\\
Crossovers among random, self-avoiding, and neighbor-avoiding walks''
}

\author{Nobu C.~Shirai}
\thanks{These authors contributed equally}
\email[Corresponding author. ]{shirai@cc.mie-u.ac.jp}
\affiliation{
Center for Information Technologies and Networks, Mie University, Tsu, Mie 514-8507, Japan
}
\author{Naoyuki Sakumichi}
\thanks{These authors contributed equally}
\email[Corresponding author. ]{sakumichi@gel.t.u-tokyo.ac.jp}
\affiliation{
Faculty of Social Informatics, ZEN University, Shinjuku, Zushi, Kanagawa 249-0007, Japan
}
\affiliation{
Department of Chemistry and Biotechnology, The University of Tokyo, Bunkyo-ku, Tokyo 113-8656, Japan
}

\date{\today}

\maketitle

\section{Tables of enumeration results for the random walk with on-axis constraint} Tables~\ref{tab:W_RW_n_m_s_n1to3} to \ref{tab:W_RW_n_m_s_n20} present complete lists of $W^\mathrm{RW}_n(r)$ and $W^\mathrm{RW}_{n,m}(r)$ for $n=1,\dots,20$ and $0\leq r \leq n$.
We do not ignore the overlap between $\omega(0)$ and $\omega(n)$ for $W^\mathrm{RW}_n(0)$.
We also provide lists of $W^\mathrm{RW}_{n,m}(n-8)$ and $W^\mathrm{RW}_{n,m}(n-10)$ in Tables~\ref{tab:W_RW_n_m_s8_n21to23_m0to9} to \ref{tab:W_RW_n_m_s10_n21to26_m21to30}. 
The results in Secs.~\MainSecPoly{}, \MainSecNEE{}, and \MainSecScaling{} of the main text can be reproduced using the data provided in these tables.
For $W^\mathrm{RW}_n(r)\geq 1$ (i.e., when $\omega$ exists), $r$ must be odd for odd values of $n$ and even for even values of $n$ (see Table~\ref{tab:W_RW_n_s_n1to20}).

\begin{table}[H]
\centering
\caption{\label{tab:W_RW_n_m_s_n1to3}List of $W^\mathrm{RW}_{n,m}(n-s)$ for $n=1,2,3$.}
\begin{tabular}{rrrr}
\hline
\hline
$n$ & $s$ & $m$ & $W$\\
\hline
1 & 0 & 0 & 1\\
\hline
2 & 0 & 0 & 1\\
2 & 2 & 1 & 6\\
\hline
3 & 0 & 0 & 1\\
3 & 2 & 0 & 4\\
3 & 2 & 1 & 10\\
3 & 2 & 2 & 1\\
\hline
\hline
\end{tabular}
\end{table}

\begin{table}[H]
\centering
\caption{\label{tab:W_RW_n_m_s_n4}List of $W^{\mathrm{RW}}_{n,m}(n-s)$ for $n=4$.}
\begin{tabular}{@{}*{5}{r} @{}}
\hline
\hline
\headercell{\\$m$} & & \multicolumn{3}{c@{}}{$s$}\\
\cmidrule(l){3-5}
 & & $0$ & $2$ & $4$\\
\hline
$0$ & & $1$ & $12$ & $0$\\
$1$ & & $0$ & $14$ & $24$\\
$2$ & & $0$ & $2$ & $30$\\
$3$ & & $0$ & $0$ & $30$\\
$4$ & & $0$ & $0$ & $6$\\
\hline
\hline
\end{tabular}
\end{table}

\begin{table}[H]
\centering
\caption{\label{tab:W_RW_n_m_s_n5}List of $W^{\mathrm{RW}}_{n,m}(n-s)$ for $n=5$.}
\begin{tabular}{@{}*{5}{r} @{}}
\hline
\hline
\headercell{\\$m$} & & \multicolumn{3}{c@{}}{$s$}\\
\cmidrule(l){3-5}
 & & $0$ & $2$ & $4$\\
\hline
$0$ & & $1$ & $24$ & $44$\\
$1$ & & $0$ & $18$ & $96$\\
$2$ & & $0$ & $3$ & $99$\\
$3$ & & $0$ & $0$ & $40$\\
$4$ & & $0$ & $0$ & $30$\\
$5$ & & $0$ & $0$ & $0$\\
$6$ & & $0$ & $0$ & $1$\\
\hline
\hline
\end{tabular}
\end{table}

\begin{table}[H]
\centering
\caption{\label{tab:W_RW_n_m_s_n6}List of $W^{\mathrm{RW}}_{n,m}(n-s)$ for $n=6$.}
\begin{tabular}{@{}*{6}{r} @{}}
\hline
\hline
\headercell{\\$m$} & & \multicolumn{4}{c@{}}{$s$}\\
\cmidrule(l){3-6}
 & & $0$ & $2$ & $4$ & $6$\\
\hline
$0$ & & $1$ & $40$ & $184$ & $0$\\
$1$ & & $0$ & $22$ & $288$ & $264$\\
$2$ & & $0$ & $4$ & $207$ & $384$\\
$3$ & & $0$ & $0$ & $62$ & $390$\\
$4$ & & $0$ & $0$ & $50$ & $516$\\
$5$ & & $0$ & $0$ & $2$ & $30$\\
$6$ & & $0$ & $0$ & $2$ & $180$\\
$7$ & & $0$ & $0$ & $0$ & $90$\\
$8$ & & $0$ & $0$ & $0$ & $0$\\
$9$ & & $0$ & $0$ & $0$ & $6$\\
\hline
\hline
\end{tabular}
\end{table}

\begin{table}[H]
\centering
\caption{\label{tab:W_RW_n_m_s_n7}List of $W^{\mathrm{RW}}_{n,m}(n-s)$ for $n=7$.}
\begin{tabular}{@{}*{6}{r} @{}}
\hline
\hline
\headercell{\\$m$} & & \multicolumn{4}{c@{}}{$s$}\\
\cmidrule(l){3-6}
 & & $0$ & $2$ & $4$ & $6$\\
\hline
$0$ & & $1$ & $60$ & $516$ & $552$\\
$1$ & & $0$ & $26$ & $648$ & $1472$\\
$2$ & & $0$ & $5$ & $367$ & $1904$\\
$3$ & & $0$ & $0$ & $92$ & $1124$\\
$4$ & & $0$ & $0$ & $71$ & $1476$\\
$5$ & & $0$ & $0$ & $4$ & $264$\\
$6$ & & $0$ & $0$ & $3$ & $352$\\
$7$ & & $0$ & $0$ & $0$ & $240$\\
$8$ & & $0$ & $0$ & $0$ & $30$\\
$9$ & & $0$ & $0$ & $0$ & $40$\\
$10$ & & $0$ & $0$ & $0$ & $0$\\
$11$ & & $0$ & $0$ & $0$ & $0$\\
$12$ & & $0$ & $0$ & $0$ & $1$\\
\hline
\hline
\end{tabular}
\end{table}

\begin{table}[H]
\centering
\caption{\label{tab:W_RW_n_m_s_n8}List of $W^{\mathrm{RW}}_{n,m}(n-s)$ for $n=8$.}
\begin{tabular}{@{}*{7}{r} @{}}
\hline
\hline
\headercell{\\$m$} & & \multicolumn{5}{c@{}}{$s$}\\
\cmidrule(l){3-7}
 & & $0$ & $2$ & $4$ & $6$ & $8$\\
\hline
$0$ & & $1$ & $84$ & $1172$ & $2616$ & $0$\\
$1$ & & $0$ & $30$ & $1224$ & $6128$ & $3312$\\
$2$ & & $0$ & $6$ & $591$ & $6120$ & $6624$\\
$3$ & & $0$ & $0$ & $130$ & $3204$ & $7920$\\
$4$ & & $0$ & $0$ & $93$ & $3038$ & $8262$\\
$5$ & & $0$ & $0$ & $6$ & $816$ & $6570$\\
$6$ & & $0$ & $0$ & $4$ & $542$ & $3438$\\
$7$ & & $0$ & $0$ & $0$ & $398$ & $4740$\\
$8$ & & $0$ & $0$ & $0$ & $78$ & $954$\\
$9$ & & $0$ & $0$ & $0$ & $68$ & $1404$\\
$10$ & & $0$ & $0$ & $0$ & $6$ & $480$\\
$11$ & & $0$ & $0$ & $0$ & $0$ & $720$\\
$12$ & & $0$ & $0$ & $0$ & $2$ & $180$\\
$13$ & & $0$ & $0$ & $0$ & $0$ & $120$\\
$14$ & & $0$ & $0$ & $0$ & $0$ & $0$\\
$15$ & & $0$ & $0$ & $0$ & $0$ & $0$\\
$16$ & & $0$ & $0$ & $0$ & $0$ & $6$\\
\hline
\hline
\end{tabular}
\end{table}

\begin{table}[H]
\centering
\caption{\label{tab:W_RW_n_m_s_n9}List of $W^{\mathrm{RW}}_{n,m}(n-s)$ for $n=9$.}
\begin{tabular}{@{}*{7}{r} @{}}
\hline
\hline
\headercell{\\$m$} & & \multicolumn{5}{c@{}}{$s$}\\
\cmidrule(l){3-7}
 & & $0$ & $2$ & $4$ & $6$ & $8$\\
\hline
$0$ & & $1$ & $112$ & $2320$ & $8936$ & $8040$\\
$1$ & & $0$ & $34$ & $2064$ & $18624$ & $23760$\\
$2$ & & $0$ & $7$ & $891$ & $15684$ & $38060$\\
$3$ & & $0$ & $0$ & $176$ & $7236$ & $31920$\\
$4$ & & $0$ & $0$ & $116$ & $5661$ & $32525$\\
$5$ & & $0$ & $0$ & $8$ & $1612$ & $23840$\\
$6$ & & $0$ & $0$ & $5$ & $836$ & $11150$\\
$7$ & & $0$ & $0$ & $0$ & $556$ & $12340$\\
$8$ & & $0$ & $0$ & $0$ & $132$ & $5425$\\
$9$ & & $0$ & $0$ & $0$ & $96$ & $4120$\\
$10$ & & $0$ & $0$ & $0$ & $12$ & $1480$\\
$11$ & & $0$ & $0$ & $0$ & $0$ & $1200$\\
$12$ & & $0$ & $0$ & $0$ & $3$ & $1015$\\
$13$ & & $0$ & $0$ & $0$ & $0$ & $400$\\
$14$ & & $0$ & $0$ & $0$ & $0$ & $100$\\
$15$ & & $0$ & $0$ & $0$ & $0$ & $0$\\
$16$ & & $0$ & $0$ & $0$ & $0$ & $50$\\
$17$ & & $0$ & $0$ & $0$ & $0$ & $0$\\
$18$ & & $0$ & $0$ & $0$ & $0$ & $0$\\
$19$ & & $0$ & $0$ & $0$ & $0$ & $0$\\
$20$ & & $0$ & $0$ & $0$ & $0$ & $1$\\
\hline
\hline
\end{tabular}
\end{table}

\begin{table}[H]
\centering
\caption{\label{tab:W_RW_n_m_s_n10}List of $W^{\mathrm{RW}}_{n,m}(n-s)$ for $n=10$.}
\begin{tabular}{@{}*{8}{r} @{}}
\hline
\hline
\headercell{\\$m$} & & \multicolumn{6}{c@{}}{$s$}\\
\cmidrule(l){3-8}
 & & $0$ & $2$ & $4$ & $6$ & $8$ & $10$\\
\hline
$0$ & & $1$ & $144$ & $4164$ & $25180$ & $40744$ & $0$\\
$1$ & & $0$ & $38$ & $3216$ & $46280$ & $113136$ & $48240$\\
$2$ & & $0$ & $8$ & $1279$ & $34244$ & $157008$ & $114048$\\
$3$ & & $0$ & $0$ & $230$ & $14124$ & $118728$ & $165528$\\
$4$ & & $0$ & $0$ & $140$ & $9704$ & $100129$ & $177600$\\
$5$ & & $0$ & $0$ & $10$ & $2694$ & $65554$ & $180942$\\
$6$ & & $0$ & $0$ & $6$ & $1234$ & $29346$ & $116088$\\
$7$ & & $0$ & $0$ & $0$ & $726$ & $27078$ & $125502$\\
$8$ & & $0$ & $0$ & $0$ & $188$ & $12587$ & $78348$\\
$9$ & & $0$ & $0$ & $0$ & $124$ & $7598$ & $51474$\\
$10$ & & $0$ & $0$ & $0$ & $18$ & $3756$ & $40812$\\
$11$ & & $0$ & $0$ & $0$ & $0$ & $1606$ & $20850$\\
$12$ & & $0$ & $0$ & $0$ & $4$ & $1802$ & $22872$\\
$13$ & & $0$ & $0$ & $0$ & $0$ & $740$ & $14418$\\
$14$ & & $0$ & $0$ & $0$ & $0$ & $276$ & $4896$\\
$15$ & & $0$ & $0$ & $0$ & $0$ & $0$ & $720$\\
$16$ & & $0$ & $0$ & $0$ & $0$ & $92$ & $6432$\\
$17$ & & $0$ & $0$ & $0$ & $0$ & $8$ & $2010$\\
$18$ & & $0$ & $0$ & $0$ & $0$ & $0$ & $1200$\\
$19$ & & $0$ & $0$ & $0$ & $0$ & $0$ & $300$\\
$20$ & & $0$ & $0$ & $0$ & $0$ & $2$ & $120$\\
$21$ & & $0$ & $0$ & $0$ & $0$ & $0$ & $150$\\
$22$ & & $0$ & $0$ & $0$ & $0$ & $0$ & $0$\\
$23$ & & $0$ & $0$ & $0$ & $0$ & $0$ & $0$\\
$24$ & & $0$ & $0$ & $0$ & $0$ & $0$ & $0$\\
$25$ & & $0$ & $0$ & $0$ & $0$ & $0$ & $6$\\
\hline
\hline
\end{tabular}
\end{table}

\begin{table}[H]
\centering
\caption{\label{tab:W_RW_n_m_s_n11}List of $W^{\mathrm{RW}}_{n,m}(n-s)$ for $n=11$.}
\begin{tabular}{@{}*{8}{r} @{}}
\hline
\hline
\headercell{\\$m$} & & \multicolumn{6}{c@{}}{$s$}\\
\cmidrule(l){3-8}
 & & $0$ & $2$ & $4$ & $6$ & $8$ & $10$\\
\hline
$0$ & & $1$ & $180$ & $6944$ & $61624$ & $154752$ & $127016$\\
$1$ & & $0$ & $42$ & $4728$ & $100040$ & $409352$ & $417792$\\
$2$ & & $0$ & $9$ & $1767$ & $66684$ & $512432$ & $752040$\\
$3$ & & $0$ & $0$ & $292$ & $24964$ & $347136$ & $806272$\\
$4$ & & $0$ & $0$ & $165$ & $15551$ & $261621$ & $807240$\\
$5$ & & $0$ & $0$ & $12$ & $4098$ & $148470$ & $747630$\\
$6$ & & $0$ & $0$ & $7$ & $1749$ & $66686$ & $492348$\\
$7$ & & $0$ & $0$ & $0$ & $908$ & $50530$ & $397980$\\
$8$ & & $0$ & $0$ & $0$ & $246$ & $23950$ & $309564$\\
$9$ & & $0$ & $0$ & $0$ & $152$ & $12882$ & $172590$\\
$10$ & & $0$ & $0$ & $0$ & $24$ & $6440$ & $161304$\\
$11$ & & $0$ & $0$ & $0$ & $0$ & $2218$ & $47520$\\
$12$ & & $0$ & $0$ & $0$ & $5$ & $2623$ & $73948$\\
$13$ & & $0$ & $0$ & $0$ & $0$ & $1082$ & $44568$\\
$14$ & & $0$ & $0$ & $0$ & $0$ & $463$ & $29115$\\
$15$ & & $0$ & $0$ & $0$ & $0$ & $0$ & $240$\\
$16$ & & $0$ & $0$ & $0$ & $0$ & $134$ & $14016$\\
$17$ & & $0$ & $0$ & $0$ & $0$ & $16$ & $7884$\\
$18$ & & $0$ & $0$ & $0$ & $0$ & $0$ & $3000$\\
$19$ & & $0$ & $0$ & $0$ & $0$ & $0$ & $2400$\\
$20$ & & $0$ & $0$ & $0$ & $0$ & $3$ & $648$\\
$21$ & & $0$ & $0$ & $0$ & $0$ & $0$ & $700$\\
$22$ & & $0$ & $0$ & $0$ & $0$ & $0$ & $150$\\
$23$ & & $0$ & $0$ & $0$ & $0$ & $0$ & $0$\\
$24$ & & $0$ & $0$ & $0$ & $0$ & $0$ & $0$\\
$25$ & & $0$ & $0$ & $0$ & $0$ & $0$ & $60$\\
$26$ & & $0$ & $0$ & $0$ & $0$ & $0$ & $0$\\
$27$ & & $0$ & $0$ & $0$ & $0$ & $0$ & $0$\\
$28$ & & $0$ & $0$ & $0$ & $0$ & $0$ & $0$\\
$29$ & & $0$ & $0$ & $0$ & $0$ & $0$ & $0$\\
$30$ & & $0$ & $0$ & $0$ & $0$ & $0$ & $1$\\
\hline
\hline
\end{tabular}
\end{table}

\begin{table}[H]
\centering
\caption{\label{tab:W_RW_n_m_s_n12}List of $W^{\mathrm{RW}}_{n,m}(n-s)$ for $n=12$.}
\begin{tabular}{@{}*{9}{r} @{}}
\hline
\hline
\headercell{\\$m$} & & \multicolumn{7}{c@{}}{$s$}\\
\cmidrule(l){3-9}
 & & $0$ & $2$ & $4$ & $6$ & $8$ & $10$ & $12$\\
\hline
$0$ & & $1$ & $220$ & $10936$ & $135232$ & $497812$ & $673040$ & $0$\\
$1$ & & $0$ & $46$ & $6648$ & $195192$ & $1232184$ & $2135936$ & $762096$\\
$2$ & & $0$ & $10$ & $2367$ & $119304$ & $1396420$ & $3523952$ & $2088960$\\
$3$ & & $0$ & $0$ & $362$ & $41044$ & $859856$ & $3508336$ & $3425952$\\
$4$ & & $0$ & $0$ & $191$ & $23610$ & $591901$ & $3157036$ & $4045536$\\
$5$ & & $0$ & $0$ & $14$ & $5860$ & $298914$ & $2592266$ & $4474608$\\
\hline
\hline
\end{tabular}
\end{table}

\begin{widetext}

\begin{table}[H]
\centering
\caption{\label{tab:W_RW_n_m_s_n13}List of $W^{\mathrm{RW}}_{n,m}(n-s)$ for $n=13$.}
\begin{tabular}{@{}*{9}{r} @{}}
\hline
\hline
\headercell{\\$m$} & & \multicolumn{7}{c@{}}{$s$}\\
\cmidrule(l){3-9}
 & & $0$ & $2$ & $4$ & $6$ & $8$ & $10$ & $12$\\
\hline
$0$ & & $1$ & $264$ & $16452$ & $272136$ & $1408808$ & $2729100$ & $2112320$\\
$1$ & & $0$ & $50$ & $9024$ & $352088$ & $3225080$ & $8523536$ & $7716240$\\
$2$ & & $0$ & $11$ & $3091$ & $200000$ & $3327452$ & $13320524$ & $15314572$\\
$3$ & & $0$ & $0$ & $440$ & $63844$ & $1885056$ & $12326000$ & $19182856$\\
$4$ & & $0$ & $0$ & $218$ & $34313$ & $1201125$ & $10223036$ & $20593720$\\
$5$ & & $0$ & $0$ & $16$ & $8016$ & $550710$ & $7493046$ & $20493536$\\
\hline
\hline
\end{tabular}
\end{table}

\begin{table}[H]
\centering
\caption{\label{tab:W_RW_n_m_s_n14}List of $W^{\mathrm{RW}}_{n,m}(n-s)$ for $n=14$.}
\begin{tabular}{@{}*{10}{r} @{}}
\hline
\hline
\headercell{\\$m$} & & \multicolumn{8}{c@{}}{$s$}\\
\cmidrule(l){3-10}
 & & $0$ & $2$ & $4$ & $6$ & $8$ & $10$ & $12$ & $14$\\
\hline
$0$ & & $1$ & $312$ & $23840$ & $510516$ & $3587520$ & $9562412$ & $11563632$ & $0$\\
$1$ & & $0$ & $54$ & $11904$ & $596864$ & $7559648$ & $28970984$ & $41263208$ & $12673920$\\
$2$ & & $0$ & $12$ & $3951$ & $318444$ & $7155476$ & $42585272$ & $77326744$ & $39683520$\\
$3$ & & $0$ & $0$ & $526$ & $95036$ & $3765072$ & $36632712$ & $91852488$ & $72247800$\\
$4$ & & $0$ & $0$ & $246$ & $48116$ & $2242761$ & $28170600$ & $91812492$ & $93803616$\\
$5$ & & $0$ & $0$ & $18$ & $10602$ & $946410$ & $18832602$ & $84269360$ & $109430136$\\
\hline
\hline
\end{tabular}
\end{table}

\begin{table}[H]
\centering
\caption{\label{tab:W_RW_n_m_s_n15}List of $W^{\mathrm{RW}}_{n,m}(n-s)$ for $n=15$.}
\begin{tabular}{@{}*{10}{r} @{}}
\hline
\hline
\headercell{\\$m$} & & \multicolumn{8}{c@{}}{$s$}\\
\cmidrule(l){3-10}
 & & $0$ & $2$ & $4$ & $6$ & $8$ & $10$ & $12$ & $14$\\
\hline
$0$ & & $1$ & $364$ & $33484$ & $903880$ & $8362224$ & $29916784$ & $49090288$ & $36484128$\\
$1$ & & $0$ & $58$ & $15336$ & $962160$ & $16213256$ & $86681136$ & $174909328$ & $146793728$\\
$2$ & & $0$ & $13$ & $4959$ & $486264$ & $14193348$ & $119193660$ & $319383312$ & $318838624$\\
$3$ & & $0$ & $0$ & $620$ & $136484$ & $6989328$ & $95477856$ & $363339224$ & $449612480$\\
$4$ & & $0$ & $0$ & $275$ & $65499$ & $3922409$ & $68576336$ & $342805268$ & $522873104$\\
$5$ & & $0$ & $0$ & $20$ & $13654$ & $1538134$ & $42332350$ & $291764000$ & $549408256$\\
\hline
\hline
\end{tabular}
\end{table}

 \begin{longtable}{@{}*{11}{r}@{}}
\caption{\label{tab:W_RW_n_m_s_n16}List of $W^{\mathrm{RW}}_{n,m}(n-s)$ for $n=16$.}\\
\hline
\hline
\endfirsthead

\caption{\textit{(Continued)}}\\
\hline
\hline
\headercell{\\$m$} & & \multicolumn{9}{c@{}}{$s$}\\
\cmidrule(l){3-11}
 & & $0$ & $2$ & $4$ & $6$ & $8$ & $10$ & $12$ & $14$ & $16$\\
\hline
\endhead

\hline
\multicolumn{11}{r}{\textit{(Table continued)}}\\
\endfoot

\hline
\hline
\endlastfoot

\headercell{\\$m$} & & \multicolumn{9}{c@{}}{$s$}\\
\cmidrule(l){3-11}
 & & $0$ & $2$ & $4$ & $6$ & $8$ & $10$ & $12$ & $14$ & $16$\\
\hline
$0$ & & $1$ & $420$ & $45804$ & $1524744$ & $18088032$ & $84977400$ & $182639672$ & $204708128$ & $0$\\
$1$ & & $0$ & $62$ & $19368$ & $1487840$ & $32330952$ & $233192720$ & $643451960$ & $810450960$ & $218904768$\\
$2$ & & $0$ & $14$ & $6127$ & $717224$ & $26377392$ & $299583796$ & $1136263528$ & $1690085936$ & $770667072$\\
$3$ & & $0$ & $0$ & $722$ & $190244$ & $12231576$ & $224272560$ & $1232002032$ & $2291509672$ & $1544869104$\\
$4$ & & $0$ & $0$ & $305$ & $86966$ & $6507161$ & $151475028$ & $1096106732$ & $2530486396$ & $2192376096$\\
$5$ & & $0$ & $0$ & $22$ & $17208$ & $2388410$ & $87082266$ & $870628448$ & $2514334416$ & $2702599440$\\
\end{longtable}

 \begin{longtable}{@{}*{11}{r}@{}}
\caption{\label{tab:W_RW_n_m_s_n17}List of $W^{\mathrm{RW}}_{n,m}(n-s)$ for $n=17$.}\\
\hline
\hline
\endfirsthead

\caption{\textit{(Continued)}}\\
\hline
\hline
\headercell{\\$m$} & & \multicolumn{9}{c@{}}{$s$}\\
\cmidrule(l){3-11}
 & & $0$ & $2$ & $4$ & $6$ & $8$ & $10$ & $12$ & $14$ & $16$\\
\hline
\endhead

\hline
\multicolumn{11}{r}{\textit{(Table continued)}} \\
\endfoot

\hline
\hline
\endlastfoot

\headercell{\\$m$} & & \multicolumn{9}{c@{}}{$s$}\\
\cmidrule(l){3-11}
 & & $0$ & $2$ & $4$ & $6$ & $8$ & $10$ & $12$ & $14$ & $16$\\
\hline
$0$ & & $1$ & $480$ & $61256$ & $2468712$ & $36713772$ & $221870344$ & $614123820$ & $898975960$ & $648529392$\\
$1$ & & $0$ & $66$ & $24048$ & $2221712$ & $60682904$ & $573473456$ & $2113467112$ & $3581485496$ & $2852506944$\\
$2$ & & $0$ & $15$ & $7467$ & $1027404$ & $46459880$ & $689336204$ & $3576910708$ & $7387428532$ & $6732082044$\\
$3$ & & $0$ & $0$ & $832$ & $258564$ & $20391456$ & $484417312$ & $3678809288$ & $9784285184$ & $10482385896$\\
$4$ & & $0$ & $0$ & $336$ & $113045$ & $10335501$ & $309478632$ & $3090022344$ & $10396808352$ & $13174348428$\\
$5$ & & $0$ & $0$ & $24$ & $21300$ & $3571014$ & $166788006$ & $2300295816$ & $9819509304$ & $14656653720$\\
\end{longtable}

 \begin{longtable}{@{}*{12}{r}@{}}
\caption{\label{tab:W_RW_n_m_s_n18}List of $W^{\mathrm{RW}}_{n,m}(n-s)$ for $n=18$.}\\
\hline
\hline
\endfirsthead

\caption{\textit{(Continued)}}\\
\hline
\hline
\headercell{\\$m$} & & \multicolumn{9}{c@{}}{$s$}\\
\cmidrule(l){3-12}
 & & $0$ & $2$ & $4$ & $6$ & $8$ & $10$ & $12$ & $14$ & $16$ & $18$\\
\hline
\endhead

\hline
\multicolumn{12}{r}{\textit{(Table continued)}} \\
\endfoot

\hline
\hline
\endlastfoot

\headercell{\\$m$} & & \multicolumn{10}{c@{}}{$s$}\\
\cmidrule(l){3-12}
 & & $0$ & $2$ & $4$ & $6$ & $8$ & $10$ & $12$ & $14$ & $16$ & $18$\\
\hline
$0$ & & $1$ & $544$ & $80332$ & $3858956$ & $70560108$ & $538054084$ & $1892115152$ & $3496496500$ & $3709636344$ & $0$\\
$1$ & & $0$ & $70$ & $29424$ & $3220248$ & $108228560$ & $1306625776$ & $6296694592$ & $13923889376$ & $16129670984$ & $3891176352$\\
$2$ & & $0$ & $16$ & $8991$ & $1435380$ & $78236232$ & $1473849664$ & $10152026316$ & $28248831968$ & $36930355032$ & $15213370368$\\
$3$ & & $0$ & $0$ & $950$ & $343884$ & $32640376$ & $976637960$ & $9886008440$ & $36365623056$ & $55798965576$ & $33318054168$\\
$4$ & & $0$ & $0$ & $368$ & $144288$ & $15827745$ & $593149656$ & $7859587928$ & $37065112980$ & $67433864900$ & $51372030720$\\
$5$ & & $0$ & $0$ & $26$ & $25966$ & $5171810$ & $301273010$ & $5507254288$ & $33278998968$ & $71929595848$ & $67075347024$\\
\hline
\end{longtable}

\newpage

\begin{longtable}{@{}*{12}{r}@{}}
\caption{\label{tab:W_RW_n_m_s_n19}List of $W^{\mathrm{RW}}_{n,m}(n-s)$ for $n=19$.}\\
\hline
\hline
\endfirsthead

\caption{\textit{(Continued)}}\\
\hline
\hline
\headercell{\\$m$} & & \multicolumn{10}{c@{}}{$s$}\\
\cmidrule(l){3-12}
 & & $0$ & $2$ & $4$ & $6$ & $8$ & $10$ & $12$ & $14$ & $16$ & $18$\\
\hline
\endhead

\hline
\multicolumn{12}{r}{\textit{(Table continued)}} \\
\endfoot

\hline
\hline
\endlastfoot

\headercell{\\$m$} & & \multicolumn{10}{c@{}}{$s$}\\
\cmidrule(l){3-12}
 & & $0$ & $2$ & $4$ & $6$ & $8$ & $10$ & $12$ & $14$ & $16$ & $18$\\
\hline
$0$ & & $1$ & $612$ & $103560$ & $5851096$ & $129356500$ & $1222907256$ & $5391815300$ & $12409673296$ & $16720159740$ & $11790401800$\\
$1$ & & $0$ & $74$ & $35544$ & $4549304$ & $184798728$ & $2788663064$ & $17223979880$ & $48912541480$ & $73478136256$ & $56317980960$\\
$2$ & & $0$ & $17$ & $10711$ & $1962404$ & $126809736$ & $2962288192$ & $26373392076$ & $96765361824$ & $167874721784$ & $143543987720$\\
$3$ & & $0$ & $0$ & $1076$ & $448836$ & $50471712$ & $1858880112$ & $24325240360$ & $120224838384$ & $250762432336$ & $243761533440$\\
$4$ & & $0$ & $0$ & $401$ & $181271$ & $23497021$ & $1077834060$ & $18363186740$ & $117201428404$ & $295777112088$ & $329468108680$\\
$5$ & & $0$ & $0$ & $28$ & $31242$ & $7289590$ & $518255926$ & $12161878904$ & $100002714560$ & $304786671032$ & $388378413600$\\
\end{longtable}

\newpage

\begin{longtable}{@{}*{13}{r}@{}}
\caption{\label{tab:W_RW_n_m_s_n20}List of $W^{\mathrm{RW}}_{n,m}(n-s)$ for $n=20$.}\\
\hline
\hline
\endfirsthead

\caption{\textit{(Continued)}}\\
\hline
\hline
\headercell{\\$m$} & & \multicolumn{11}{c@{}}{$s$}\\
\cmidrule(l){3-13}
 & & $0$ & $2$ & $4$ & $6$ & $8$ & $10$ & $12$ & $14$ & $16$ & $18$ & $20$\\
\hline
\endhead

\hline
\multicolumn{13}{r}{\textit{(Table continued)}}\\
\endfoot

\hline
\hline
\endlastfoot

\headercell{\\$m$} & & \multicolumn{11}{c@{}}{$s$}\\
\cmidrule(l){3-13}
 & & $0$ & $2$ & $4$ & $6$ & $8$ & $10$ & $12$ & $14$ & $16$ & $18$ & $20$\\
\hline
$0$ & & $1$ & $684$ & $131504$ & $8638480$ & $227589504$ & $2625286352$ & $14322531084$ & $40683432604$ & $67289840344$ & $68496828560$ & $0$\\
$1$ & & $0$ & $78$ & $42456$ & $6284840$ & $303906776$ & $5625406136$ & $43706172200$ & $157278964944$ & $297500478672$ & $324427841936$ & $70742410800$\\
$2$ & & $0$ & $18$ & $12639$ & $2632584$ & $198896588$ & $5648544020$ & $63499792684$ & $301218432500$ & $675578496268$ & $807820681848$ & $304117097184$\\
$3$ & & $0$ & $0$ & $1210$ & $576244$ & $75755328$ & $3369701616$ & $55563554760$ & $359387321928$ & $993899418376$ & $1339516647416$ & $722801929632$\\
$4$ & & $0$ & $0$ & $435$ & $224594$ & $33960789$ & $1872169836$ & $39961462284$ & $334721858252$ & $1140618177428$ & $1756088565724$ & $1204098700560$\\
$5$ & & $0$ & $0$ & $30$ & $37164$ & $10036914$ & $855424074$ & $25111433136$ & $271669308824$ & $1133544756096$ & $2002354855416$ & $1665604794816$\\
\end{longtable}

\begin{table}[H]
\centering
\caption{\label{tab:W_RW_n_m_s8_n21to23_m0to9}List of $W^\mathrm{RW}_{n,m}(n-8)$ for $n=21,22,23$ and $0 \leq m \leq 9$.}
\begin{tabular}{@{}*{12}{r} @{}}
\hline
\hline
\headercell{\\$n$} & \multicolumn{10}{c@{}}{$m$}\\
\cmidrule(l){3-12}
 & & $0$ & $1$ & $2$ & $3$ & $4$ & $5$ & $6$ & $7$ & $8$ & $9$\\
\hline
$21$ & & $386219812$ & $483700152$ & $303174052$ & $110796416$ & $47952901$ & $13540950$ & $4827481$ & $1653310$ & $575240$ & $248262$\\
$22$ & & $634830332$ & $748063424$ & $450674540$ & $158398656$ & $66336201$ & $17944314$ & $6306733$ & $2057898$ & $700373$ & $298068$\\
$23$ & & $1014272508$ & $1127884040$ & $655228412$ & $221931696$ & $90115665$ & $23405910$ & $8120456$ & $2530402$ & $843637$ & $354114$\\
\hline
\hline
\end{tabular}
\end{table}

\begin{table}[H]
\centering
\caption{\label{tab:W_RW_n_m_s8_n21to23_m10to20}List of $W^\mathrm{RW}_{n,m}(n-8)$ for $n=21,22,23$ and $10 \leq m \leq 20$.}
\begin{tabular}{@{}*{10}{r} @{}}
\hline
\hline
\headercell{\\$n$} & \multicolumn{8}{c@{}}{$m$}\\
\cmidrule(l){3-10}
 & & $10$ & $11$ & $12$ & $13$ & $14$ & $16$ & $17$ & $20$\\
\hline
$21$ & & $80355$ & $14358$ & $16168$ & $4942$ & $2443$ & $554$ & $96$ & $13$\\
$22$ & & $94132$ & $16166$ & $18320$ & $5372$ & $2652$ & $596$ & $104$ & $14$\\
$23$ & & $109370$ & $18082$ & $20665$ & $5810$ & $2863$ & $638$ & $112$ & $15$\\
\hline
\hline
\end{tabular}
\end{table}

\begin{table}[H]
\centering
\caption{\label{tab:W_RW_n_m_s10_n21to26_m0to8}List of $W^\mathrm{RW}_{n,m}(n-10)$ for $n=21,\dots,26$ and $0 \leq m \leq 8$.}
\begin{tabular}{@{}*{11}{r} @{}}
\hline
\hline
\headercell{\\$n$} & \multicolumn{9}{c@{}}{$m$}\\
\cmidrule(l){3-11}
 & & $0$ & $1$ & $2$ & $3$ & $4$ & $5$ & $6$ & $7$ & $8$\\
\hline
$21$ & & $5358881828$ & $10805317880$ & $10293575336$ & $5858192400$ & $3128498628$ & $1362823670$ & $556630115$ & $239828576$ & $105149195$\\
$22$ & & $10460905212$ & $19884038576$ & $18034135544$ & $9821540552$ & $5055418952$ & $2105587530$ & $837076324$ & $346782502$ & $147067286$\\
$23$ & & $19624056512$ & $35234505416$ & $30523612904$ & $15951453120$ & $7932730776$ & $3167020974$ & $1228118069$ & $490144280$ & $201560468$\\
$24$ & & $35526914104$ & $60378837144$ & $50111445612$ & $25190729712$ & $12129032420$ & $4652066650$ & $1763011954$ & $679005574$ & $271342492$\\
$25$ & & $62293921824$ & $100420638536$ & $80068367680$ & $38801376576$ & $18122241936$ & $6691168998$ & $2482338471$ & $924016176$ & $359518285$\\
$26$ & & $106123083700$ & $162599029640$ & $124865575376$ & $58445738440$ & $26523326328$ & $9444559074$ & $3435106960$ & $1237545566$ & $469616750$\\
\hline
\hline
\end{tabular}
\end{table}

\begin{table}[H]
\centering
\caption{\label{tab:W_RW_n_m_s10_n21to26_m9to20}List of $W^\mathrm{RW}_{n,m}(n-10)$ for $n=21,\dots,26$ and $9 \leq m \leq 20$.}
\begin{tabular}{@{}*{14}{r} @{}}
\hline
\hline
\headercell{\\$n$} & \multicolumn{12}{c@{}}{$m$}\\
\cmidrule(l){3-14}
 & & $9$ & $10$ & $11$ & $12$ & $13$ & $14$ & $15$ & $16$ & $17$ & $18$ & $19$ & $20$\\
\hline
$21$ & & $46213474$ & $20403756$ & $6230854$ & $4422506$ & $2156324$ & $1138921$ & $162534$ & $245658$ & $114676$ & $32107$ & $25508$ & $11440$\\
$22$ & & $63509122$ & $27137896$ & $8019556$ & $5652054$ & $2665716$ & $1395876$ & $191806$ & $296396$ & $131772$ & $35870$ & $27900$ & $13120$\\
$23$ & & $85594844$ & $35495597$ & $10163282$ & $7123980$ & $3253854$ & $1690390$ & $223554$ & $354167$ & $150424$ & $39783$ & $30308$ & $14956$\\
$24$ & & $113396186$ & $45743870$ & $12706218$ & $8870752$ & $3927494$ & $2025376$ & $257802$ & $419478$ & $170728$ & $43846$ & $32732$ & $16960$\\
$25$ & & $147950040$ & $58176237$ & $15695198$ & $10927806$ & $4693584$ & $2403795$ & $294574$ & $492836$ & $192780$ & $48059$ & $35172$ & $19144$\\
$26$ & & $190410904$ & $73113932$ & $19179704$ & $13333726$ & $5559264$ & $2828656$ & $333894$ & $574748$ & $216676$ & $52422$ & $37628$ & $21520$\\
\hline
\hline
\end{tabular}
\end{table}

\begin{table}[H]
\centering
\caption{\label{tab:W_RW_n_m_s10_n21to26_m21to30}List of $W^\mathrm{RW}_{n,m}(n-10)$ for $n=21,\dots,26$ and $21 \leq m \leq 30$.}
\begin{tabular}{@{}*{8}{r} @{}}
\hline
\hline
\headercell{\\$n$} & \multicolumn{6}{c@{}}{$m$}\\
\cmidrule(l){3-8}
 & & $21$ & $22$ & $24$ & $25$ & $26$ & $30$\\
\hline
$21$ & & $7472$ & $3184$ & $200$ & $500$ & $100$ & $11$\\
$22$ & & $8190$ & $3500$ & $220$ & $544$ & $110$ & $12$\\
$23$ & & $8916$ & $3818$ & $240$ & $588$ & $120$ & $13$\\
$24$ & & $9650$ & $4138$ & $260$ & $632$ & $130$ & $14$\\
$25$ & & $10392$ & $4460$ & $280$ & $676$ & $140$ & $15$\\
$26$ & & $11142$ & $4784$ & $300$ & $720$ & $150$ & $16$\\
\hline
\hline
\end{tabular}
\end{table}

\begin{table}[H]
\centering
\caption{\label{tab:W_RW_n_s_n1to20}List of $W^{\mathrm{RW}}_n(n-s)$ for $n=1,\dots,20$.}
\begin{tabular}{@{}*{13}{r} @{}}
\hline
\hline
\headercell{\\$n$} & & \multicolumn{11}{c@{}}{$s$}\\
\cmidrule(l){3-13}
 & & $0$ & $2$ & $4$ & $6$ & $8$ & $10$ & $12$ & $14$ & $16$ & $18$ & $20$\\
\hline
$1$ & & $1$ & $0$ & $0$ & $0$ & $0$ & $0$ & $0$ & $0$ & $0$ & $0$ & $0$\\
$2$ & & $1$ & $6$ & $0$ & $0$ & $0$ & $0$ & $0$ & $0$ & $0$ & $0$ & $0$\\
$3$ & & $1$ & $15$ & $0$ & $0$ & $0$ & $0$ & $0$ & $0$ & $0$ & $0$ & $0$\\
$4$ & & $1$ & $28$ & $90$ & $0$ & $0$ & $0$ & $0$ & $0$ & $0$ & $0$ & $0$\\
$5$ & & $1$ & $45$ & $310$ & $0$ & $0$ & $0$ & $0$ & $0$ & $0$ & $0$ & $0$\\
$6$ & & $1$ & $66$ & $795$ & $1860$ & $0$ & $0$ & $0$ & $0$ & $0$ & $0$ & $0$\\
\hline
\hline
\end{tabular}
\end{table}

\end{widetext}

\section{Tables of enumeration results for the random walk with off-axis constraints} 

Tables~\ref{tab:W_RW_n16_18_20_rvec1_1_0} to \ref{tab:W_RW_n16_18_20_rvec2_2_2} present enumeration results for the random walk under off‐axis constraints with the end-to-end vectors $\mathbf{r}=(1,1,0)$, $(2,2,0)$, $(3,3,0)$, $(1,1,1)$, and $(2,2,2)$.

\begin{longtable}{@{}*{5}{r}@{}}
\caption{List of $W^{\mathrm{RW}}_{n,m}$ for $n=16$, $18$, and $20$ with $\mathbf{r}=(1,1,0)$.} \label{tab:W_RW_n16_18_20_rvec1_1_0}\\
\hline
\hline
\endfirsthead
\caption{\textit{(Continued)}}\\
\hline
\hline
\headercell{\\$m$} & & \multicolumn{3}{c@{}}{$n$}\\
\cmidrule(l){3-5}
 & & $16$ & $18$ & $20$\\
\hline
\endhead
\hline
\multicolumn{5}{r}{\textit{(Table continued)}}\\
\endfoot
\hline
\hline
\endlastfoot
\headercell{\\$m$} & & \multicolumn{3}{c@{}}{$n$}\\
\cmidrule(l){3-5}
 & & $16$ & $18$ & $20$\\
\hline
$0$ &  & $190191620$ & $3428140888$ & $63026844160$\\
$1$ &  & $769988464$ & $15178844720$ & $303094668792$\\
$2$ &  & $1658863212$ & $35652849676$ & $770602976064$\\
$3$ &  & $2350458228$ & $55702017064$ & $1312241485224$\\
$4$ &  & $2713265976$ & $69754103152$ & $1770901993228$\\
$5$ &  & $2818871724$ & $77120599232$ & $2080061062716$\\
\end{longtable}
 
\begin{longtable}{@{}*{5}{r}@{}}
\caption{List of $W^{\mathrm{RW}}_{n,m}$ for $n=16$, $18$, and $20$ with $\mathbf{r}=(2,2,0)$.} \label{tab:W_RW_n16_18_20_rvec2_2_0}\\
\hline
\hline
\endfirsthead
\caption{\textit{(Continued)}}\\
\hline
\hline
\headercell{\\$m$} & & \multicolumn{3}{c@{}}{$n$}\\
\cmidrule(l){3-5}
 & & $16$ & $18$ & $20$\\
\hline
\endhead
\hline
\multicolumn{5}{r}{\textit{(Table continued)}}\\
\endfoot
\hline
\hline
\endlastfoot
\headercell{\\$m$} & & \multicolumn{3}{c@{}}{$n$}\\
\cmidrule(l){3-5}
 & & $16$ & $18$ & $20$\\
\hline
$0$ &  & $214321134$ & $3936256708$ & $73437085684$\\
$1$ &  & $817005292$ & $16610161564$ & $339406634316$\\
$2$ &  & $1613876652$ & $36497400900$ & $818027762598$\\
$3$ &  & $2030877836$ & $52199802128$ & $1300253046628$\\
$4$ &  & $2073396558$ & $59384754816$ & $1625375664628$\\
$5$ &  & $1895254036$ & $59450215128$ & $1762131076508$\\
\end{longtable}
 
\begin{longtable}{@{}*{5}{r}@{}}
\caption{List of $W^{\mathrm{RW}}_{n,m}$ for $n=16$, $18$, and $20$ with $\mathbf{r}=(3,3,0)$.} \label{tab:W_RW_n16_18_20_rvec3_3_0}\\
\hline
\hline
\endfirsthead
\caption{\textit{(Continued)}}\\
\hline
\hline
\headercell{\\$m$} & & \multicolumn{3}{c@{}}{$n$}\\
\cmidrule(l){3-5}
 & & $16$ & $18$ & $20$\\
\hline
\endhead
\hline
\multicolumn{5}{r}{\textit{(Table continued)}}\\
\endfoot
\hline
\hline
\endlastfoot
\headercell{\\$m$} & & \multicolumn{3}{c@{}}{$n$}\\
\cmidrule(l){3-5}
 & & $16$ & $18$ & $20$\\
\hline
$0$ &  & $174521080$ & $3376394124$ & $65489915260$\\
$1$ &  & $603124880$ & $13256708924$ & $286419857288$\\
$2$ &  & $1036297268$ & $26383954144$ & $641217487768$\\
$3$ &  & $1077392264$ & $33071149864$ & $925723786556$\\
$4$ &  & $909708764$ & $32637830796$ & $1038935074896$\\
$5$ &  & $676335920$ & $28205440160$ & $1006457868336$\\
\end{longtable}
 
\begin{longtable}{@{}*{5}{r}@{}}
\caption{List of $W^{\mathrm{RW}}_{n,m}$ for $n=15$, $17$, and $19$ with $\mathbf{r}=(1,1,1)$.} \label{tab:W_RW_n15_17_19_rvec1_1_1}\\
\hline
\hline
\endfirsthead
\caption{\textit{(Continued)}}\\
\hline
\hline
\headercell{\\$m$} & & \multicolumn{3}{c@{}}{$n$}\\
\cmidrule(l){3-5}
 & & $15$ & $17$ & $19$\\
\hline
\endhead
\hline
\multicolumn{5}{r}{\textit{(Table continued)}}\\
\endfoot
\hline
\hline
\endlastfoot
\headercell{\\$m$} & & \multicolumn{3}{c@{}}{$n$}\\
\cmidrule(l){3-5}
 & & $15$ & $17$ & $19$\\
\hline
$0$ &  & $48802530$ & $871579794$ & $15906099318$\\
$1$ &  & $185330112$ & $3642778080$ & $72576078864$\\
$2$ &  & $372621276$ & $8053393056$ & $174781584636$\\
$3$ &  & $484757328$ & $11721364464$ & $279967608120$\\
$4$ &  & $520915464$ & $13781423010$ & $357158567808$\\
$5$ &  & $507879192$ & $14407233540$ & $398857843056$\\
\end{longtable}
 
\begin{longtable}{@{}*{5}{r}@{}}
\caption{List of $W^{\mathrm{RW}}_{n,m}$ for $n=16$, $18$, and $20$ with $\mathbf{r}=(2,2,2)$.} \label{tab:W_RW_n16_18_20_rvec2_2_2}\\
\hline
\hline
\endfirsthead
\caption{\textit{(Continued)}}\\
\hline
\hline
\headercell{\\$m$} & & \multicolumn{3}{c@{}}{$n$}\\
\cmidrule(l){3-5}
 & & $16$ & $18$ & $20$\\
\hline
\endhead
\hline
\multicolumn{5}{r}{\textit{(Table continued)}}\\
\endfoot
\hline
\hline
\endlastfoot
\headercell{\\$m$} & & \multicolumn{3}{c@{}}{$n$}\\
\cmidrule(l){3-5}
 & & $16$ & $18$ & $20$\\
\hline
$0$ &  & $203654148$ & $3807068664$ & $71982759168$\\
$1$ &  & $748906992$ & $15637712220$ & $325779837624$\\
$2$ &  & $1404842772$ & $33099227766$ & $763258713120$\\
$3$ &  & $1645704648$ & $45018824832$ & $1168922375964$\\
$4$ &  & $1558615464$ & $48438912720$ & $1400578207380$\\
$5$ &  & $1313349768$ & $45732141792$ & $1451658866340$\\
\end{longtable}
 
\section{Tables of enumeration results for the self-avoiding walk with off-axis constraints} Tables~\ref{tab:W_SAW_n16_18_20_rvec1_1_0} to \ref{tab:W_SAW_n16_18_20_rvec2_2_2} present enumeration results for the self-avoiding walk under off‐axis constraints with the end-to-end vectors $\mathbf{r}=(1,1,0)$, $(2,2,0)$, $(3,3,0)$, $(1,1,1)$, and $(2,2,2)$.

\begin{longtable}{@{}*{5}{r}@{}}
\caption{List of $W^{\mathrm{SAW}}_{n,m}$ for $n=16$, $18$, and $20$ with $\mathbf{r}=(1,1,0)$.} \label{tab:W_SAW_n16_18_20_rvec1_1_0}\\
\hline
\hline
\endfirsthead
\caption{\textit{(Continued)}}\\
\hline
\hline
\headercell{\\$m$} & & \multicolumn{3}{c@{}}{$n$}\\
\cmidrule(l){3-5}
 & & $16$ & $18$ & $20$\\
\hline
\endhead
\hline
\multicolumn{5}{r}{\textit{(Table continued)}}\\
\endfoot
\hline
\hline
\endlastfoot
\headercell{\\$m$} & & \multicolumn{3}{c@{}}{$n$}\\
\cmidrule(l){3-5}
 & & $16$ & $18$ & $20$\\
\hline
$0$ &  & $5166272$ & $69837676$ & $961626196$\\
$1$ &  & $14554248$ & $210434872$ & $3092844936$\\
$2$ &  & $23634640$ & $361796614$ & $5626150472$\\
$3$ &  & $29101516$ & $469085764$ & $7668628936$\\
$4$ &  & $30160560$ & $509302254$ & $8714709368$\\
$5$ &  & $27530788$ & $486409348$ & $8700976904$\\
$6$ &  & $21946968$ & $421306694$ & $7855833224$\\
$7$ &  & $17200052$ & $328940000$ & $6546282568$\\
$8$ &  & $11749888$ & $243922966$ & $5026650268$\\
$9$ &  & $5668832$ & $166054308$ & $3631289652$\\
$10$ &  & $2512928$ & $91772260$ & $2448200852$\\
$11$ &  & $702240$ & $44758856$ & $1449017712$\\
$12$ &  & $245568$ & $17022460$ & $767049268$\\
$13$ &  & $12168$ & $6024472$ & $345100976$\\
$14$ &  & $4952$ & $1120984$ & $136470892$\\
$15$ &  & $0$ & $341432$ & $42424336$\\
$16$ &  & $0$ & $9928$ & $11638720$\\
$17$ &  & $0$ & $0$ & $1528960$\\
$18$ &  & $0$ & $0$ & $419920$\\
\end{longtable}
 
\begin{longtable}{@{}*{5}{r}@{}}
\caption{List of $W^{\mathrm{SAW}}_{n,m}$ for $n=16$, $18$, and $20$ with $\mathbf{r}=(2,2,0)$.} \label{tab:W_SAW_n16_18_20_rvec2_2_0}\\
\hline
\hline
\endfirsthead
\caption{\textit{(Continued)}}\\
\hline
\hline
\headercell{\\$m$} & & \multicolumn{3}{c@{}}{$n$}\\
\cmidrule(l){3-5}
 & & $16$ & $18$ & $20$\\
\hline
\endhead
\hline
\multicolumn{5}{r}{\textit{(Table continued)}}\\
\endfoot
\hline
\hline
\endlastfoot
\headercell{\\$m$} & & \multicolumn{3}{c@{}}{$n$}\\
\cmidrule(l){3-5}
 & & $16$ & $18$ & $20$\\
\hline
$0$ &  & $10646214$ & $143119552$ & $1965526230$\\
$1$ &  & $25190732$ & $367908624$ & $5463192792$\\
$2$ &  & $35164752$ & $551295168$ & $8755645744$\\
$3$ &  & $38321840$ & $639826244$ & $10768743868$\\
$4$ &  & $35116666$ & $627334496$ & $11160954832$\\
$5$ &  & $28618456$ & $539935708$ & $10205167388$\\
$6$ &  & $20073806$ & $421407128$ & $8427284668$\\
$7$ &  & $11758112$ & $294727968$ & $6404255384$\\
$8$ &  & $5887280$ & $180146316$ & $4464880114$\\
$9$ &  & $2404744$ & $97289892$ & $2805278152$\\
$10$ &  & $851152$ & $45650752$ & $1595806128$\\
$11$ &  & $227520$ & $18912260$ & $819722744$\\
$12$ &  & $53540$ & $6298008$ & $377157800$\\
$13$ &  & $5384$ & $1951544$ & $150103944$\\
$14$ &  & $936$ & $370816$ & $54038728$\\
$15$ &  & $0$ & $77472$ & $14857880$\\
$16$ &  & $0$ & $4760$ & $3809460$\\
$17$ &  & $0$ & $0$ & $549832$\\
$18$ &  & $0$ & $0$ & $109996$\\
\end{longtable}
 
\begin{longtable}{@{}*{5}{r}@{}}
\caption{List of $W^{\mathrm{SAW}}_{n,m}$ for $n=16$, $18$, and $20$ with $\mathbf{r}=(3,3,0)$.} \label{tab:W_SAW_n16_18_20_rvec3_3_0}\\
\hline
\hline
\endfirsthead
\caption{\textit{(Continued)}}\\
\hline
\hline
\headercell{\\$m$} & & \multicolumn{3}{c@{}}{$n$}\\
\cmidrule(l){3-5}
 & & $16$ & $18$ & $20$\\
\hline
\endhead
\hline
\multicolumn{5}{r}{\textit{(Table continued)}}\\
\endfoot
\hline
\hline
\endlastfoot
\headercell{\\$m$} & & \multicolumn{3}{c@{}}{$n$}\\
\cmidrule(l){3-5}
 & & $16$ & $18$ & $20$\\
\hline
$0$ &  & $12145828$ & $164181166$ & $2264470516$\\
$1$ &  & $27668964$ & $410016256$ & $6154531856$\\
$2$ &  & $36209352$ & $586578178$ & $9520784968$\\
$3$ &  & $35961976$ & $638802060$ & $11171592136$\\
$4$ &  & $29321104$ & $574113214$ & $10889376916$\\
$5$ &  & $18541648$ & $443633756$ & $9184393272$\\
$6$ &  & $9181576$ & $286707364$ & $6876361092$\\
$7$ &  & $3791364$ & $155916644$ & $4522176232$\\
$8$ &  & $1291600$ & $73153050$ & $2606803420$\\
$9$ &  & $327012$ & $29622596$ & $1330149800$\\
$10$ &  & $68912$ & $10036208$ & $606845304$\\
$11$ &  & $11744$ & $2866096$ & $243915424$\\
$12$ &  & $0$ & $663176$ & $85535760$\\
$13$ &  & $0$ & $89992$ & $25522544$\\
$14$ &  & $0$ & $14368$ & $6145020$\\
$15$ &  & $0$ & $0$ & $1163592$\\
$16$ &  & $0$ & $0$ & $130616$\\
$17$ &  & $0$ & $0$ & $16792$\\
\end{longtable}
 
\begin{longtable}{@{}*{5}{r}@{}}
\caption{List of $W^{\mathrm{SAW}}_{n,m}$ for $n=15$, $17$, and $19$ with $\mathbf{r}=(1,1,1)$.} \label{tab:W_SAW_n15_17_19_rvec1_1_1}\\
\hline
\hline
\endfirsthead
\caption{\textit{(Continued)}}\\
\hline
\hline
\headercell{\\$m$} & & \multicolumn{3}{c@{}}{$n$}\\
\cmidrule(l){3-5}
 & & $15$ & $17$ & $19$\\
\hline
\endhead
\hline
\multicolumn{5}{r}{\textit{(Table continued)}}\\
\endfoot
\hline
\hline
\endlastfoot
\headercell{\\$m$} & & \multicolumn{3}{c@{}}{$n$}\\
\cmidrule(l){3-5}
 & & $15$ & $17$ & $19$\\
\hline
$0$ &  & $2065224$ & $27600294$ & $376708614$\\
$1$ &  & $4981020$ & $71808168$ & $1054932582$\\
$2$ &  & $7146390$ & $109819674$ & $1717387650$\\
$3$ &  & $8096802$ & $131214066$ & $2160978120$\\
$4$ &  & $7907406$ & $133594146$ & $2307981432$\\
$5$ &  & $6524862$ & $121449624$ & $2184759510$\\
$6$ &  & $5338638$ & $97599108$ & $1888444020$\\
$7$ &  & $3720960$ & $74750190$ & $1486684080$\\
$8$ &  & $1884750$ & $51927378$ & $1100967288$\\
$9$ &  & $806940$ & $29488758$ & $757045074$\\
$10$ &  & $237768$ & $14243730$ & $455235666$\\
$11$ &  & $84408$ & $5670300$ & $242730888$\\
$12$ &  & $4584$ & $1880142$ & $110997684$\\
$13$ &  & $2778$ & $417048$ & $42689466$\\
$14$ &  & $0$ & $114816$ & $14370096$\\
$15$ &  & $0$ & $0$ & $3491472$\\
$16$ &  & $0$ & $2352$ & $531624$\\
$17$ &  & $0$ & $0$ & $164052$\\
\end{longtable}
 
\begin{longtable}{@{}*{5}{r}@{}}
\caption{List of $W^{\mathrm{SAW}}_{n,m}$ for $n=16$, $18$, and $20$ with $\mathbf{r}=(2,2,2)$.} \label{tab:W_SAW_n16_18_20_rvec2_2_2}\\
\hline
\hline
\endfirsthead
\caption{\textit{(Continued)}}\\
\hline
\hline
\headercell{\\$m$} & & \multicolumn{3}{c@{}}{$n$}\\
\cmidrule(l){3-5}
 & & $16$ & $18$ & $20$\\
\hline
\endhead
\hline
\multicolumn{5}{r}{\textit{(Table continued)}}\\
\endfoot
\hline
\hline
\endlastfoot
\headercell{\\$m$} & & \multicolumn{3}{c@{}}{$n$}\\
\cmidrule(l){3-5}
 & & $16$ & $18$ & $20$\\
\hline
$0$ &  & $11601924$ & $156091548$ & $2145259332$\\
$1$ &  & $27141576$ & $398093292$ & $5927801076$\\
$2$ &  & $36971328$ & $585683244$ & $9367310820$\\
$3$ &  & $38967768$ & $662777556$ & $11297342460$\\
$4$ &  & $34166112$ & $628920120$ & $11419745052$\\
$5$ &  & $26332764$ & $518530164$ & $10117394196$\\
$6$ &  & $15794652$ & $384343392$ & $8028047988$\\
$7$ &  & $7810836$ & $241689936$ & $5823716700$\\
$8$ &  & $3307308$ & $130545708$ & $3761061096$\\
$9$ &  & $1157484$ & $62070624$ & $2149833732$\\
$10$ &  & $316548$ & $25791702$ & $1110040464$\\
$11$ &  & $75720$ & $9069840$ & $516763176$\\
$12$ &  & $9132$ & $2708838$ & $211832040$\\
$13$ &  & $996$ & $628944$ & $76039092$\\
$14$ &  & $0$ & $113352$ & $23505324$\\
$15$ &  & $0$ & $8520$ & $5759712$\\
$16$ &  & $0$ & $1884$ & $1098084$\\
$17$ &  & $0$ & $0$ & $191832$\\
$18$ &  & $0$ & $0$ & $16992$\\
\end{longtable}